\documentclass[prx,aps,superscriptaddress,nofootinbib,twocolumn,notitlepage,floatfix,10pt]{revtex4-2}

\usepackage{amsmath,mathtools,amsthm,amssymb,pifont}
\usepackage[utf8]{inputenc}
\usepackage[american]{babel}
\usepackage{graphicx,xcolor,bbold,titlesec}
\usepackage{braket}
\usepackage{MnSymbol}
\newtheorem{theorem}{Theorem}

\usepackage[colorlinks,
bookmarksopen,
bookmarksnumbered,
citecolor=red,
linkcolor=red,
pdfstartview=false,
urlcolor=red]{hyperref}
\usepackage{tikz,ifthen}
\usepackage{bbold}
\usepackage{tikz-network}
\usetikzlibrary{patterns,decorations.pathreplacing,calligraphy}
\usepackage{color}
\usepackage{MnSymbol}
\definecolor{myyellow}{RGB}{240,188,66}
\definecolor{myorange}{RGB}{255,102,0}
\definecolor{myorangel}{RGB}{255,204,153}
\definecolor{myblue}{RGB}{66,135,245}

\renewcommand{\boxed}[1]{%
  \framebox{\raisebox{0pt}[0.4\baselineskip][0.025\baselineskip]{\hbox to 0.25cm{\hss#1\hss}}}}

 \newcommand{\titleinfo}{Magic spreading in random quantum circuits}
 
\begin{document}
 \title{\titleinfo} 

\author{ 
Xhek Turkeshi}
\affiliation{Institut f\"ur Theoretische Physik, Universit\"at zu K\"oln, Z\"ulpicher Strasse 77, 50937 K\"oln, Germany}

\author{Emanuele Tirrito }
\affiliation{The Abdus Salam International Centre for Theoretical Physics (ICTP), Strada Costiera 11, 34151 Trieste, Italy}
\affiliation{Pitaevskii BEC Center, CNR-INO and Dipartimento di Fisica,
Università di Trento, Via Sommarive 14, Trento, I-38123, Italy}

\author{Piotr Sierant}
\affiliation{ICFO-Institut de Ciències Fotòniques, The Barcelona Institute of Science and Technology, Av. Carl Friedrich Gauss 3, 08860 Castelldefels (Barcelona), Spain}

\date{\today}

\begin{abstract}
\textit{Magic} is the resource that quantifies the amount of beyond-Clifford operations necessary for universal quantum computing. 
It bounds the cost of classically simulating quantum systems via stabilizer circuits central to quantum error correction and computation. How rapidly do generic many-body dynamics generate magic resources under the constraints of locality and unitarity? We address this central question by exploring magic spreading in brick-wall random unitary circuits. 
We explore scalable magic measures intimately connected to the algebraic structure of the Clifford group.
These metrics enable the investigation of the spreading of magic for system sizes of up to $N=1024$ qudits, surpassing the previous state-of-the-art, which was restricted to about a dozen qudits. 
We demonstrate that magic resources equilibrate on timescales logarithmic in the system size, 
akin to anti-concentration and Hilbert space delocalization phenomena, but qualitatively different from the spreading of entanglement entropy. 
As random circuits are minimal models for chaotic dynamics, we conjecture that our findings describe the phenomenology of magic resources growth in a broad class of chaotic many-body systems.
\end{abstract}
\maketitle

Quantum computers require several types of resources to solve computational tasks faster than classical computers~\cite{shor1997polynomialtimealgorithmsfor,chitambar2019quantumresourcetheories}. 
Entanglement is one such resource, but alone, it is insufficient to guarantee that a quantum computer outperforms its classical counterpart. Indeed, stabilizer states can attain extensive entanglement under Clifford operations while being efficiently simulatable on classical computers via the Gottesman-Knill theorem~\cite{gottesman1998heisenbergrep,gottesmann1998faulttolerantquantum,aaronson2004improvedsimulationof}.
nonstabilizerness, colloquially called ``magic", quantifies the additional non-Clifford operations required to perform a given quantum operation, constituting another necessary ingredient for the quantum speedup~\cite{bravyi2005universalquantumcomputation}. 
Understanding how magic resources build up and propagate in many-body quantum systems emerges as a fundamental question, with potential impact on current and near-term quantum devices~\cite{liu2022manybodyquantummagic}.

The question of magic resources generation is ambitious but challenging. Until a few years ago, measures of magic required minimization procedures over large spaces, resulting in prohibitive computational costs for even a few qubits~\cite{Veitch2014theresourcetheory}. Recently, mana and stabilizer entropies have been introduced as scalable measures of magic~\cite{gross2006hudsonstheoremfor,leone2022stabilizerrenyientropy,leone2023nonstabilizerness,leone2024stabilizerentropiesmonotonesmagicstate, Gu24pseudo}. Subsequent developments in tensor network methods~\cite{haug2023quantifyingnonstabilizernessof,haug2023stabilizerentropiesand,lami2023nonstabilizernessviaperfect,tarabunga2024nonstabilizernessmatrixproductstates} and Monte-Carlo approaches~\cite{tarabunga2023manybodymagic,Tarabunga2024magicingeneralized} 
have provided a powerful toolbox to characterize the nonstabilizerness of ground states~\cite{frau2024nonstabilizernessversusentanglementmatrix} while enabling hybrid Clifford-tensor network algorithms~\cite{mello2024hybridstabilizermatrixproduct,qian2024augmentingdensitymatrixrenormalization, masotllima2024stabilizertensornetworks,lami2024quantumstatedesignsclifford,lami2024learningstabilizergroupmatrix}. 
Despite these successes, the time evolution of magic resources in many-body systems remains a largely open question. 
The rapidly growing entanglement limits the traditional tensor network methods and brute-force exact simulations to small system sizes. With these limitations, Ref.~\cite{rattacaso2023stabilizer} concluded that a quantum quench in an integrable system results in a linear growth of nonstabilizerness over time, similar to entanglement entropy~\cite{nahum2017quantum}.

\begin{figure}
    \centering
    \includegraphics[width=1\linewidth]{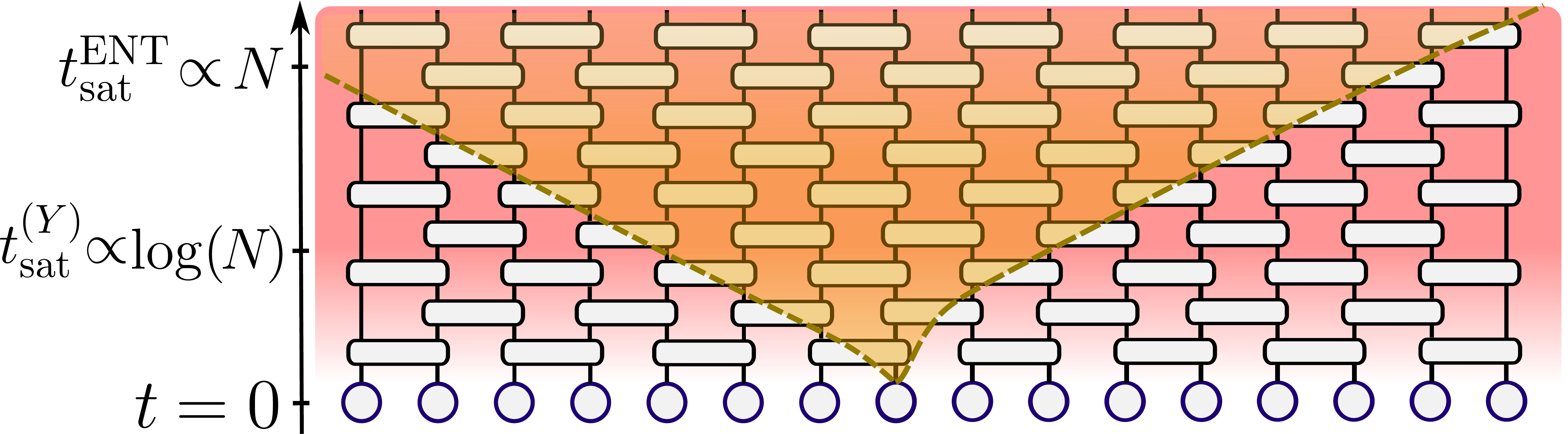}
    \caption{A system of $N$ qudits is prepared at time $t=0$ in a product state $\ket{\Psi_0}$ with low-magic resources. Evolution under a quantum circuit comprising local Haar-random gates increases the nonstabilizerness of the state (denoted by the red gradient) and scrambles quantum information (symbolized by the light-cone). The nonstabilizerness approaches its long-time saturation value up to a given tolerance $\epsilon \ll 1$ at time $t^{\mathrm{mag}}_{\mathrm{sat}} \propto \ln N$, scaling logarithmically with $N$, while distant qudits become entangled only after a longer time $t^{\mathrm{ent}}_{\mathrm{sat}}\propto N$.
    }
    \label{fig:haard_circuit}
\end{figure}

This work investigates the magic spreading under generic, non-integrable, local unitary quantum dynamics. To that end, we focus on Haar random brick-wall circuits of qudits. 
Unambiguous identification of the features of the growth of magic resources requires access to large system sizes. 
For this reason, inspired by the algebraic structure of the Clifford group, we consider the family of generalized stabilizer entropies (GSE). The latter constitute good measures of nonstabilizerness for many-body systems and include stabilizer Rényi entropy (SRE) as a particular example. 
We combine the replica trick and Haar average methods to express the circuit-averaged generalized stabilizer entropies as tensor network contractions. 
In particular, the GSE can be expressed as a $k=3$ replica quantity for qutrits while necessitating $k=4$ replica for qubits. In both cases, the replica tensor network 
methods allow us to investigate systems of up to $N\le 1024$ qubits. 
Our main finding is that the long-time saturation value of GSE entropy is reached, up to a tolerance $\epsilon \ll 1$, at times $t^{\mathrm{mag}}_{\mathrm{sat}} \propto \ln N$, scaling logarithmically with system size, see Fig.~\ref{fig:haard_circuit}.

\section{Generalized stabilizer entropies}
\label{sec:CSS}
We start by discussing the GSE, characterizing the magic resources for systems of qudits~\cite{leone2022stabilizerrenyientropy, leonetoappear}. 
A finite field with $d$ elements is denoted by $\mathbb{Z}_d=\{0,1,\dots,d-1\}$, $\mathcal{H}_d=\mathrm{span}[\{|m\rangle\}_{m\in \mathbb{Z}_d}]$ is the local Hilbert space dimension of a $d$-dimensional qudit, and $N$ is the number of qudits. 
The GSEs are tied to the algebraic structure of the Pauli~\cite{knill1996nonbinary} and Clifford groups~\cite{gottesman1998heisenbergrep} acting on $\mathcal{H}^{\otimes N}_d$. 
The Pauli operators $X$ and $Z$ are defined as 
\begin{equation}
    X = \sum_{m=0}^{d-1} |m\rangle \langle m\oplus_d 1|\;,\qquad  Z = \sum_{m=0}^{d-1} \omega^m |m\rangle\langle m|\;,
\end{equation} 
with ${a\oplus_d b = a+b \;\mathrm{mod}\; d}$ representing the sum in $\mathbb{Z}_d$ and ${\omega = e^{2\pi i/d}}$~\cite{gottesman1998theoryoffaulttolerant}. The Pauli group ${\mathcal{P}}_N(d)=\{ X_1^{r_1^x} Z_1^{r_1^z} X_2^{r_2^x} Z_2^{r_2^z}\cdots X_N^{r_N^x} Z_N^{r_N^z} \;| \; r_k^\alpha \in \mathbb{Z}_d\}$ is generated by tensor products of Pauli operators (called also Pauli strings). 
The Clifford group $\mathcal{C}_{N,d}$ consists of the unitary $C$ mapping, up to a global phase, a Pauli string $P$ to a Pauli string $\omega^r P' = C P C^\dagger$ with $r\in \mathbb{Z}_d$. 
Stabilizer states are defined as $\mathrm{STAB}_{N,d}=\{ C|0\rangle^{\otimes N}\;|\; C\in \mathcal{C}_{N,d}\}$, and magic or non-stabilizer states are those not belonging to this set.

The theory of GSE is intimately connected to the structure of the commutants of the Clifford group for $k$ copies of the system~\cite{gross2021schurweylduality,montealegremora2022dualitytheoryfor}.
For a given set $\mathcal{E}$, we define its $k$-th commutant as the set of operators $W$ acting on $k$ copies of the system such that $\mathrm{Comm}_k(\mathcal{E})=\{ W \;|\; [W,E^{\otimes k}]=0\;\text{for all }E\in \mathcal{E}\}$. 
For 
the unitary group $\mathcal{U}(d^N)$, the commutant is built of the representation of permutation operators $\mathrm{Comm}_k(\mathcal{U}(d^N)) = \{ W_\pi \;|\; \pi \in S_k\}$~\cite{mele2024introductiontohaar}. 
By duality, since $\mathcal{C}_{N,d}\subset \mathcal{U}(d^N)$, it follows that $\mathrm{Comm}_k(\mathcal{U}(d^N))\subset \mathrm{Comm}_k(\mathcal{C}_{N,d})$. 
Importantly, the Clifford commutant contains  $|\mathrm{Comm}_k(\mathcal{C}_{N,d})|=\prod_{m=0}^{k-2} (d^m+1)$ elements, whereas $|\mathrm{Comm}_k(\mathcal{U}(d^N))|=k!$. These two numbers coincide for $d>2$ when $k\le2$, and for $d=2$ when $k\le 3$. 
Hence, the elements of the commutant, when applied to a state $\left( \ket{\Psi}\bra{\Psi} \right) ^{\otimes k}$ cannot distinguish whether $\ket{\Psi}$ is a stabilizer state or whether it has non-vanishing magic resources if $k \le 3$ for qubits and $k \le 2$ for $d\geq 3$.

On the other hand, when $k\ge 3$ for qudits (and $k\ge 4$ for qubits), the Clifford commutant is strictly greater than 
$\mathrm{Comm}_k(\mathcal{U}(d^N))$. 
In that case, the \textit{intrinsic Clifford commutant}~\cite{leonetoappear}, defined as $\overline{\mathrm{Comm}}_k(\mathcal{C}_{N,d})=\mathrm{Comm}_k(\mathcal{C}_{N,d})
\setminus
\mathrm{Comm}_k(\mathcal{U}(d^N))$, is a non-empty set, whose elements $W$, when applied to 
$\left( \ket{\Psi}\bra{\Psi} \right) ^{\otimes k}$ can distinguish if the state is a stabilizer state or not.
Let us define the 
GSE, $M_W$, and the associated generalized stabilizer purity $\zeta_W$ by 
\begin{equation}
    M_W\equiv -\ln[\zeta_W(|\Psi\rangle)]\;,\quad \zeta_W\equiv\mathrm{tr}(W |\Psi\rangle\langle\Psi|^{\otimes k})\; \label{eq:ydef}.
\end{equation}
For any $W\in \overline{\mathrm{Comm}}_k(\mathcal{C}_{N,d})$, the generalized stabilizer purity 
$\zeta_W(|\Psi\rangle)\le1$, with the equality holding if and only if $|\Psi\rangle$ is a stabilizer state, as we show in~\cite{supmat}.
This implies that the GSE $M_W(|\Psi\rangle)\ge 0$, with the equality holding if and only if $|\Psi\rangle$ is a stabilizer state. 
Moreover, for any operator $W$ from the intrinsic Clifford commutant, the associated GSE $M_W$ 
is a measure of magic for many-body systems: (i) $M_W(|\Psi\rangle)\ge 0$ and $M_W=0$ if and only if $|\Psi\rangle$ is a stabilizer state, (ii) $M_W(C|\Psi\rangle) = M_W(|\Psi\rangle)$ for $C\in \mathcal{C}_{N,d}$, a Clifford unitary, (iii) it is additive $M_W(|\Phi\rangle\otimes |\Psi\rangle) = M_W(|\Phi\rangle) + M_W(|\Psi\rangle)$. 
The subclass of stabilizer R\'enyi entropies are monotones for generic stabilizer protocols for qubits~\cite{leone2024stabilizerentropiesmonotonesmagicstate}. In~\cite{supmat}, we extend this reasoning showing that SRE is monotone also for qudits ($d \geq 3$), and argue that monotonicity under Pauli measurements of arbitrary GSE is not  guaranteed, presenting an explicit counterexample for qutrits.

To understand better the scope of GSE, we first consider how SRE arise from~\eqref{eq:ydef}.
At $k=4$, the intrinsic Clifford commutant contains six operators, one of which being $Q_4=\sum_{P\in \mathcal{P}_N(d)} (P\otimes P^\dagger)^{\otimes 2}/d^N$. 
This operator leads to the second SRE~\cite{leone2022stabilizerrenyientropy} $M_2\equiv M_{Q_4} = -\ln[\zeta_{Q_4}]$ with $\zeta_{Q_4} = \sum_{P\in \mathcal{P}_N(d)} |\langle \Psi|P|\Psi\rangle|^4/d^N$. 
Similarly, 
the SRE $M_\alpha$ of arbitrary integer index $\alpha\geq 2$, $M_\alpha \equiv \ln[|\langle \Psi|P|\Psi\rangle|^{2\alpha}/d^N]/(1-\alpha)$, fulfills $M_\alpha \propto M_{Q_{2\alpha}}$ with $Q_{2\alpha}=\sum_{P\in \mathcal{P}_N(d)} (P\otimes P^\dagger)^{\otimes \alpha}/d^N$. With a slight abuse of notation, throughout this text, we will consider $M_2 \equiv M_{Q_4}$. 
An example of GSE beyond the family of SREs is by the operator ${Y_d\equiv \sum_{P\in \mathcal{P}_N(d)} (P \otimes P \otimes P^{d-2})/d^N}$, which belongs to the intrinsic Clifford commutant for $k=3$ replicas in any qudit system with $d\ge3$ prime.  
Throughout this paper, by $M_Y$ we denote the GSE induced by the operator $Y_d$, with the dimension  $d $ inferred from context. 

The above examples provide concrete measures of the magic $M_W$ that require $k=3$ copies for qudits with odd prime $d$ and $k=4$ for qubits. 
A figure of merit valid for any stabilizer purity $\zeta_W$ is their efficient tensor network representation. In fact, the Clifford commutant operators $T$ acting on $N$ qudit systems, reduce to tensor products of operators $W=w^{\otimes N}$ acting on individual qudits. 
As a result, Eq.~\eqref{eq:ydef} is efficiently computable via tensor network methods, either by exact contractions~\cite{haug2023quantifyingnonstabilizernessof,tarabunga2024nonstabilizernessmatrixproductstates} or by sampling methods~\cite{lami2023nonstabilizernessviaperfect}, cf.~\cite{supmat} for details.

\section{Brick-wall Haar random quantum circuits.}
We consider a one-dimensional chain of $N$ qudits and study the spreading of the GSEs $M_W$ under unitary dynamics generated by brick-wall Haar random quantum circuits (see Fig. \ref{fig:haard_circuit}). The evolution operator of the considered brick-wall circuit reads $U_t = \prod_{r=1}^t U^{(r)}$, where $t$ is the circuit depth (also referred to as time). Numbering the qudits by $i=1,...,N$, the layers $U^{(r)}$ are fixed as
\begin{equation}
    U^{(2m)} = \prod_{i=1}^{N/2-1} U_{2i,2i+1}\;,\quad U^{(2m+1)} = \prod_{i=1}^{N/2} U_{2i-1,2i}\;,
    \label{eq:randC}
\end{equation}
comprising two-qubit gates $U_{i,j}$ chosen independently with the Haar distribution on the unitary group $\mathcal{U}(d^2)$. The initial state $\ket{\Psi_0} $ is chosen as the stabilizer state $|\Psi_0\rangle = |0\rangle^{\otimes N}$, with $M_W=0$ for any $W$ in the intrinsic Clifford commutant. 
How do the magic resources of $|\Psi_t\rangle$, quantified by the GSEs, 
increase under the dynamics of the circuit~\eqref{eq:randC}? 

The problem at hand is stochastic due to the randomness of the gates. Denoting with $\mathbb{E}(\bullet)$ the average over the circuit realizations, we consider quenched and annealed averages of the GSEs, defined respectively as $\overline{M_W} \equiv \mathbb{E}[M_W(|\Psi_t\rangle)] = -\mathbb{E}[\ln[\zeta_W(|\Psi_t\rangle)]]$ and $\tilde{M}_W \equiv -\ln[\mathbb{E}[\zeta_W(|\Psi_t\rangle)]]$. 
Exact numerical simulation of the random circuit \eqref{eq:randC} provides access to $\ket{\Psi_t}$, allowing for calculation of the quenched and the annealed averages of the GSEs and exposing the self-averaging of $\zeta_W(|\Psi_t\rangle)$. The circuit-to-circuit fluctuations of $\zeta_W(|\Psi_t\rangle)$ around its average value $\mathbb{E}[\zeta_W(|\Psi_t\rangle)]$ are suppressed with the increase of the system size $N$ and decay rapidly in time as demonstrated in Sec.~\ref{sec:methods} for several choices of $W$. The self-averaging of $\zeta_W(|\Psi_t\rangle)$ implies that $\overline{M_W}$ and $\tilde{M}_W$ 
approach each other with increase of $N$ and $t$. Therefore, 
the annealed average $\tilde{M}_W$ may be chosen to quantify the time evolution of the GSEs under the random circuits.

\section{Annealed average of generalized stabilizer  entropies.}
The calculation of the annealed average $\tilde{M}_W$ is facilitated by a replica trick and the Weingarten calculus, which allow us to map computation of $\tilde{M}_W$ for the random Haar circuits to a contraction of a two-dimensional tensor network. The latter can be efficiently computed to provide insights into the time evolution of the GSEs for systems comprising hundreds of qudits, far beyond the reach of exact simulation of the system.

For convenience, we employ the superoperator formalism~\cite{Choi1975}: $A\mapsto |A\rrangle$, $U_t A U^\dagger\mapsto (U_t\otimes U_t^\ast) |A\rrangle$, and $\llangle A|B\rrangle = \mathrm{tr}(A^\dagger B)$. In particular, we have~\footnote{As is common in the literature, we make a slight abuse of notation by implicitly reshaping $(\mathcal{H}_d^{\otimes N})^{\otimes k}\mapsto (\mathcal{H}_d^{\otimes j})^{\otimes N}$.}
\begin{equation}\zeta_W(|\Psi_t\rangle) =  \llangle W| (U_t\otimes U_t^\ast)^{\otimes k} |\rho_0^{\otimes k}\rrangle\;,
   \label{eq:replicaYd}
\end{equation}
where $\rho_0 = |\Psi_0\rangle\langle \Psi_0|$ is the initial state's density matrix. 
To calculate the average $\mathbb{E}[\zeta_W(|\Psi_t\rangle)]$ over the circuit realizations, we observe that~\eqref{eq:replicaYd} is linear in $(U_t\otimes U_t^\ast)^{\otimes k}$, implying that we can first average the superoperator corresponding to the circuit and then calculate the matrix element in~\eqref{eq:replicaYd}. The former, due to the statistical independence of the two-body gates at various spatial and temporal locations, reduces to evaluating the two qudit transfer matrix $\mathcal{T}^{(k)}_{i,i+1} \equiv \mathbb{E}_\mathrm{Haar}[(U_{i,i+1}\otimes U^\ast_{i,i+1})^{\otimes k}]$. 
These expressions are formulated in terms of the unitary commutant $\mathrm{Comm}_k(\mathcal{U}(d^N))$ which, as aforementinoed, consists of permutation operators. Up to reshaping, we express $\mathcal{T}^{(k)}$ via the corresponding permutation states $|\tau\rrangle_i$ acting on the $k$-replica qudits at sites $i, i+1$, leading to the expression
\begin{equation}
   \mathcal{T}^{(k)}_{i,i+1} = \sum_{\pi,\tau\in S_k} \mathrm{Wg}_{\tau,\sigma}(d^2) |\tau\rrangle_i |\tau\rrangle_{i+1}\llangle \sigma |_i \llangle \sigma|_{i+1}\;,
\end{equation}
where $\llangle b_1,\bar{b}_1,\dots,b_k,\bar{b}_k| \tau\rrangle = \prod_{m=1}^k \delta_{b_m,\bar{b}_{\tau(m)}}$ for each $k$-replica qudit basis state $|b_1,\bar{b}_1,\dots,b_k,\bar{b}_k\rrangle$ and $\mathrm{Wg}_{\tau,\sigma}(d^2)$ denotes the Weingarten symbol~\cite{collins2022weingartencalculus}. 
The lattice structure induced by the circuit requires contraction of $\mathcal{T}^{(k)}_{i,i+1}$ between the even and odd layers~\eqref{eq:randC}, with the overlaps $G_{\sigma,\tau}(d)\equiv \llangle \sigma |\tau\rrangle = d^{\#(\sigma^{-1}\tau)}$ taken into account, where $\#(\tau)$ denotes the number of cycles for $\tau\in S_k$. We reabsorb these overlaps by defining the tensors
\begin{equation}
\begin{split}
	\mathcal{T}^{(k)}_{i,i+1} &\equiv 
	\begin{tikzpicture}[baseline=(current  bounding  box.center), scale=1]
\draw[thick] (-1.75,-0.2) -- (-1.75,0.4);
\draw[thick] (-1.25,-0.2)-- (-1.25,0.4);
\draw[thick, fill=myblue, rounded corners=2pt] (-1.9,0.3) rectangle (-1.1,-0.1);
\end{tikzpicture}
\equiv \sum_{\pi_1,\pi_2,\pi,\tau\in S_k} \mathrm{Wg}_{\tau,\pi}(d^2) \times \\ & G_{\pi,\pi_1}(d)G_{\pi,\pi_2}(d) |\tau\rrangle_i |\tau\rrangle_{i+1}   \llangle \hat{\pi}_1|_i \llangle \hat{\pi}_2|_{i+1}\;,
\end{split}
\label{eq:wten}
\end{equation}
with the states $|\hat{\sigma}\rrangle$ satisfying $\llangle \hat{\sigma}|\tau\rrangle = \delta_{\sigma,\tau}$. 
The contraction with the first layer of unitary gates is fixed by the replica boundary condition
\begin{equation}
\begin{split}
	\begin{tikzpicture}[baseline=(current  bounding  box.center), scale=1]
\draw[thick] (-1.75,0) -- (-1.75,0.4);
\draw[thick] (-1.25,0)-- (-1.25,0.4);
\draw[thick, fill=myyellow, rounded corners=2pt] (-1.9,0.3) rectangle (-1.1,-0.1) (-1.5,0.08) node{$+$};
\end{tikzpicture}
&\equiv \mathcal{T}_{i,i+1}^{(k)} |\rho_0^{\otimes k}\rrangle =\sum_{\pi \in S_k} \frac{(d^2-1)!}{(d^2+k-1)!} |\pi\rrangle_{i}|\pi\rrangle_{i+1},
\end{split}
\label{eq:initial}
\end{equation}
while the contraction with the last layer of the circuit requires
\begin{equation}
\begin{split}
	\begin{tikzpicture}[baseline=(current  bounding  box.center), scale=1]
\draw[thick] (-1.75,-0.25) -- (-1.75,0.0);
\draw[thick] (-1.25,-0.25)-- (-1.25,0.0);
\draw[thick, fill=myorange, rounded corners=2pt] (-1.9,0.3) rectangle (-1.1,-0.1) (-1.5,0.08) node{$\mathfrak{w}$};
\end{tikzpicture} \equiv \llangle w|_i\llangle w|_{i+1} \mathcal{T}^{(k)}_{i,i+1}.
\end{split}
\label{eq:lastLayer}
\end{equation}
Summarizing, the computation of the annealed average of the GSEs reduces to evaluating the tensor contraction 
\begin{equation}
    \mathbb{E}[\zeta_W(|\Psi_t\rangle)] =  
    \begin{tikzpicture}[baseline=(current  bounding  box.center), scale=0.35]
    \draw [decorate,decoration={brace},thick] (-14,-6) -- node[left]{$t$}(-14,+.3);
  \foreach \i in {1,...,3} {
   \draw[thick, fill=myorange, rounded corners=1pt]  (-1.4-4*\i,+0.3) rectangle (1.4-4*\i,-0.7) (1.4-1.33-4*\i,-0.3) node{$\mathfrak{w}$};
   }
  \foreach \kk[evaluate=\kk as \k using 0.25*\kk] in {0} {
  \pgfmathsetmacro{\col}{ifthenelse(int(mod(\kk,2))==0,"myblue","myblue")}
  \foreach \i in {1,...,6}{
    \draw[ thick] (-1.-2*\i+\k,-0.7+\k) -- (-1.-2*\i+\k,-5+\k);
  }
  \foreach \jj[evaluate=\jj as \j using -(ceil(\jj/2)-\jj/2), evaluate=\j as \fin using 3+\j] in {1,...,3} {
    \foreach \i in {1,...,\fin}{        
      \draw[thick, fill=\col, rounded corners=1pt] (-1.4-4*\i+4*\j,+0.5-1.3*\jj-0.25) rectangle (1.4-4*\i+4*\j,-0.5+-1.3*\jj-0.25);
    }
  }
  \foreach \jj[evaluate=\jj as \j using -(ceil(\jj/2)-\jj/2), evaluate=\j as \fin using 3+\j] in {4} {
    \foreach \i in {1,...,\fin}{        
      \draw[thick, fill=myyellow, rounded corners=1pt] (-1.4-4*\i+4*\j+\k,+0.5-2*\jj+2.5) rectangle (1.4-4*\i+4*\j+\k,-0.5-2*\jj+2.5) (1.4-1.33-4*\i+4*\j+\k,-0.5-2*\jj+0.5+2.5)node{$+$};
    }
  }
} \end{tikzpicture}\;.
\label{eq:Ydtensor}
\end{equation} 
The effective ``spins", i.e., the degrees of freedom at the sites of the lattice~\eqref{eq:Ydtensor}, correspond to permutations of the $k$ replicas and hence admit $q_\mathrm{eff} = k!$ values, while the tensors $\mathcal{T}_{i,i+1}^{(k)}$ can be interpreted as non-unitary gates acting on the spins. These observations constitute the basis of our numerical approach, which allows us to compute the annealed average of the GSEs $\tilde{M}_W =-\ln[\mathbb{E}[\zeta_W(|\Psi_t\rangle)]]$ for arbitrary circuit depth $t$. 
While the above discussion applies to any intrinsic Clifford commutant operator $ W $, we will specifically analyze the stabilizer R\'enyi entropy $ M_2 $ for  $k=4$  replicas in qubit ( $d=2$ ) and qutrit ($d=3$) systems, and  $M_Y $ for  $k=3$  in qutrit systems.

\section{Deep circuit limit.}
In the deep circuit limit, for $t\gg1$, the brick-wall quantum circuits form approximate $k$-designs~\cite{Brandao2016, Haferkamp2022randomquantum}. 
In that limit, the operator $U_t$ in \eqref{eq:replicaYd} can be replaced by a global 
Haar random gate $U \in \mathcal{U}(d^N)$ 
and $\mathcal{T}_{i,i+1}^{(k)}$ in the contraction \eqref{eq:Ydtensor} are substituted by 
the global gate. This allows for analytical calculation of the $\tilde{M}_W$ of interest \cite{supmat}, yielding $M_{2}^\mathrm{Haar} \equiv -\ln[4/(2^N+3)]$ for $d=2$
~\cite{turkeshi2023paulispectrummagictypical}, while for $d=3$, we find $M_2^\mathrm{Haar}=M_Y^\mathrm{Haar} = -\ln[3/(3^N+2)]$. 
Due to the concentration of Haar measure~\cite{mele2024introductiontohaar}, the fluctuations of $M_W$ with circuit realizations are strongly suppressed with the increase of $N$. Hence, the GSEs $\tilde{M}_W$ saturate at long times $t\gg 1$ under the dynamics of random circuits to $M_{W}^\mathrm{Haar}$. 
Now, we characterize the approach of $M_{W}(|\Psi_t\rangle)$ to the saturation value $M_{W}^\mathrm{Haar}$.

\begin{figure}
    \centering
    \includegraphics[width=1\linewidth]{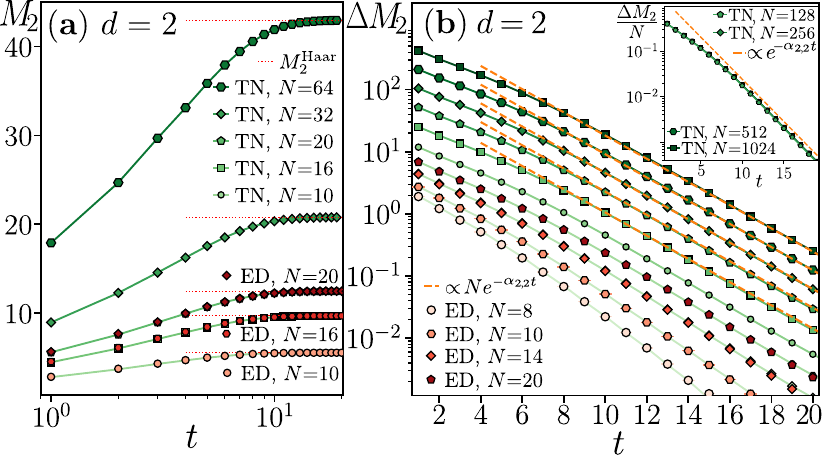}
    \caption{Time evolution of the SRE $M_2$ for $N$ qubits ($d=2$) under the brick-wall Haar random circuits. (\textbf{a}) $M_2$ abruptly saturates to $M^{\mathrm{Haar}}_2$. (\textbf{b}) The difference $\Delta M_2 = M^{\mathrm{Haar}}_2 - M_2$ approaches exponential decay $\Delta M_2 \propto N e^{-\alpha_{2,2} t}$, where $\alpha_{2,2} = 0.43(3)$, see the inset. The annealed average $\tilde{M}_2$ obtained via~\eqref{eq:Ydtensor} (denoted ``TN") and the quenched average $\overline{M}_2$ (denoted ``ED") coincide within the error bars already for $N=8$.}
    \label{fig:qubits}
\end{figure}

\section{Numerical results.}
\begin{figure}
    \centering
    \includegraphics[width=1\linewidth]{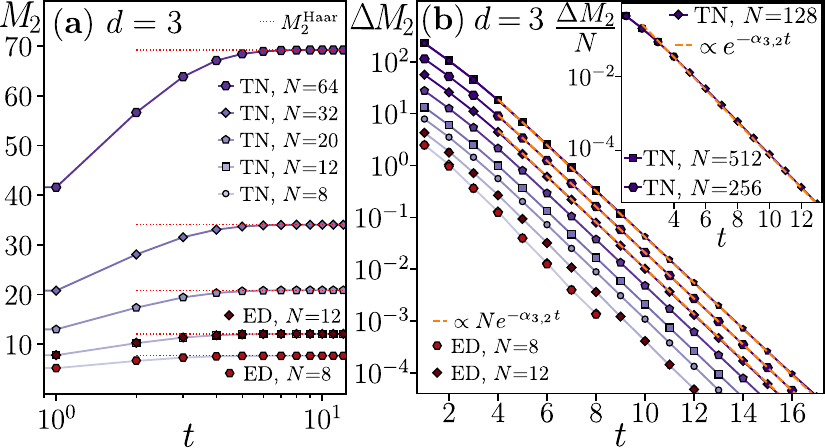}
    \caption{Dynamics of the SRE $M_2$ for $N$ qutrits ($d=3$) under random circuits. (\textbf{a}) Saturation of $M_2$ to $M^{\mathrm{Haar}}_2$ occurs similarly to the qubit case. (\textbf{b}) The difference $\Delta M_2 = M^{\mathrm{Haar}}_2 - M_2$ follows $\Delta M_2 \propto N e^{-\alpha_{3,2} t}$ with $\alpha_{3,2} = 1.03(3)$ at $t \gtrsim 5$; see the inset. The quenched $\tilde{M}_2$ and annealed $\overline{M}_2$ averages are indistinguishable on the scale of the figure for any $N$.}
    \label{fig:qutritsSRE}
\end{figure}

\begin{figure}
    \centering
    \includegraphics[width=1\linewidth]{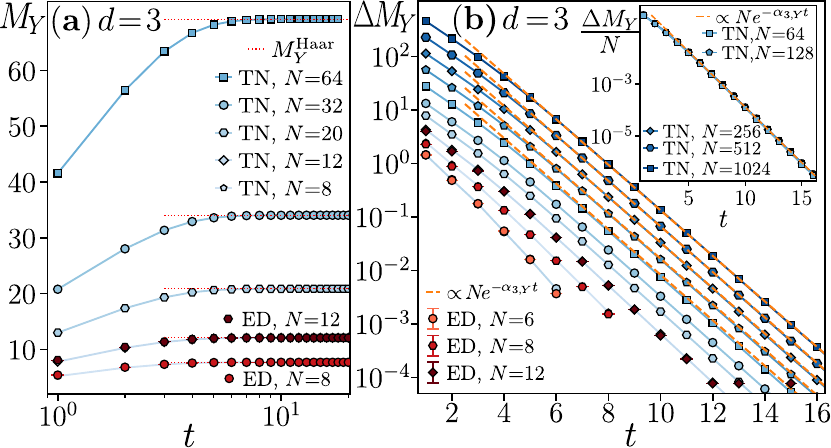}
    \caption{Dynamics of the GSE  $M_Y$ for $N$ qutrits ($d=3$) under random circuits. (\textbf{a}) Saturation of $M_Y$ to $M^{\mathrm{Haar}}_Y$ occurs similarly to the qubit case. (\textbf{b}) The difference $\Delta M_Y = M^{\mathrm{Haar}}_Y - M_Y$ follows $\Delta M_Y \propto N e^{-\alpha_{3,Y} t}$ with $\alpha_{3,Y} = 0.98(2)$ at $t \gtrsim 5$; see the inset. The quenched $\tilde{M}_Y$ and annealed $\overline{M}_Y$ averages approach each other with the increase of $N$.}
    \label{fig:qutrits}
\end{figure}

Our results for the growth of GSEs under the dynamics of random circuits are summarized in Fig.~\ref{fig:qubits} and Fig.~\ref{fig:qutritsSRE},~\ref{fig:qutrits} for qubits ($d=2$) and qutrits ($d=3$), respectively. 
We start by comparing the quenched $\tilde{M}_W$ and annealed $\overline{M}_W$ averages of the GSEs. As anticipated, already for $N=8$ qubits and qudits, we find $\tilde{M}_W \approx \overline{M}_W$, confirming 
that the quenched and annealed averages can be used interchangeably to characterize magic spreading in the considered circuits, cf. Sec.~\ref{subsec:self} for details. 
Hence, 
we focus on the annealed averages $\tilde{M}_W$ obtained from the tensor network contraction~\eqref{eq:Ydtensor}. Expressing the state of the $q_\mathrm{eff}=k!$ dimensional "spins" as a matrix product state~\cite{ran2020tensor}, 
we contract the tensor network~\eqref{eq:Ydtensor} horizontally, layer after layer. Implementing the contraction in ITensor~\cite{itensor},  
we observe that a bond dimension $\chi=\mathcal{O}(q_\mathrm{eff}^2)$ of the matrix product state is sufficient to obtain converged results, see Sec.~\ref{subsec:ten}. 
The computation requires significantly smaller resources for qutrits since $M_Y$ is a $k=3$-replica quantity resulting in $q_\mathrm{eff}=6$.  
In contrast, the $k=4$ replicas demanded to calculate $M_2$ for qubits lead to $q_\mathrm{eff}=24$, and significantly larger computational costs~\footnote{For qubits ($d=2$), this can be reduced to  $q_\mathrm{eff} = 14 $ using irreducible representations.}.
We compute $\tilde{M}_2$ for systems of up to $N=1024$ qubits ($N=512$ qutrits) with $\chi=300$ ($\chi=800$), as shown in Fig.~\ref{fig:qubits} (Fig.~\ref{fig:qutritsSRE}), and $\tilde{M}_Y$ for $N\le 1024$ qutrit systems with $\chi=300$, cf. Fig.~\ref{fig:qutrits}.

In all the cases, the GSEs $\tilde{M}_W$ are proportional to the system size $N$ already at $t=1$. Indeed, the additivity of $\tilde{M}_W$ implies that $\tilde{M}_W(\ket{\Psi_{t=1}}) = (N/2) \tilde{M}^{(2)}_W$, where $\tilde{M}^{(2)}_W$ is the average GSE generated by a single two-body gate $U_{i,i+1}$, and $N/2$ is the number of two-body gates in the first layer of the circuit. 
For $t>1$, the GSEs $\tilde{M}_W$ rapidly increase towards their saturation values $M_{W}^\mathrm{Haar}$ for both $d=2$ and $d=3$.
For circuit depths $t \gtrsim 5$, the difference $\Delta M_W(t) =M^{\mathrm{Haar}}_W - \tilde{M}_W(\ket{\Psi_t})$ is proportional to the system size $N$ and decays exponentially in time:
\begin{equation}
\Delta M_d(t) = a_{d,W} N e^{-\alpha_{d,W} t},
   \label{eq:decay}
\end{equation}
where $a_{d,W}$ and $\alpha_{d,W}$ are constants (see Figs.~\ref{fig:qubits}, ~\ref{fig:qutritsSRE} and~\ref{fig:qutrits}). The exponential relaxation of the GSEs to their long-time saturation values under the dynamics of random quantum circuits is the main result of this work. 
The saturation value of GSEs is reached, up to a fixed small accuracy $\epsilon$, i.e., $\Delta M_W = \epsilon$, at time $t^{\mathrm{mag}}_{\mathrm{sat}} = \ln(N)/\alpha_{d,W} + O(1)$, scaling logarithmically with system size $N$.

\section{ Discussion.}
The brick-wall Haar random quantum circuits combine principles of locality and unitarity of time dynamics, serving as minimal models for ergodic quantum many-body systems~\cite{fisher2023randomquantumcircuits}. Random circuits provide insights into the entanglement growth~\cite{nahum2017quantum, zhou2020entanglementmembrane, Piroli2020},
the properties of operator spreading~\cite{nahum2018operator, keyserlingks1,kheman}
and spectral correlations~\cite{Bertini18, Kos18, andrea0}. 
In particular, the magic resources in eigenstates of ergodic many-body systems share properties with states of deep random circuits~\cite{turkeshi2023paulispectrummagictypical}. Hence, we conjecture that the universal features of the GSEs growth~\eqref{eq:decay}, i.e., the exponential relaxation to the saturation value at times $t^{\mathrm{mag}}_{\mathrm{sat}} \propto \ln N$
characterize the spreading of GSEs in chaotic many-body systems.
Our conjecture is supported by the following observation based on the Suzuki-Trotter decomposition~\cite{Hatano05}. Consider quantum dynamics generated by a local ergodic quantum Hamiltonian $H = \sum_j H_{j,j+1}$. The Suzuki-Trotter formula is a key element of  algorithms computing time dynamics of many-body system, e.g., the time-evolving block decimation~\cite{Vidal07}, and allows for the following approximation of the evolution operator
\begin{equation}
    e^{-i \Delta t H} \approx \prod_{k=1}^{N/2} e^{-i \Delta t H_{2k-1,2k} }\prod_{k=1}^{N/2-1} e^{-i \Delta t H_{2k,2k+1} },
    \label{eq:ST}
\end{equation}
valid for sufficiently small $\Delta t$. Eq.~\ref{eq:ST} reproduces the structure of the brick-wall quantum circuit, with unitary gates given by $U^{H}_{k,k+1} = e^{-i \Delta t H_{k,k+1} }$. The gates $U^{H}_{k,k+1}$ for generic ergodic many-body systems do not belong to the Clifford group, and their action increases the nonstabilizerness of the state of the system. Hence, each layer of the circuit defined by \eqref{eq:ST} contains extensively many gates that increase magic resources, similar to the brick-wall Haar random quantum circuits considered in this work.

The uncovered phenomenology of the GSEs parallels, as we argue in Sec.~\ref{subsec:mana}, the growth of mana~\cite{Veitch2014theresourcetheory} under the dynamics of random quantum circuits, even though mana is a magic state resource theory monotone \textit{not belonging} to the family of GSEs. The behavior of these nonstabilizerness measures is reminiscent of time evolution of participation entropy~\cite{turkeshi2023measuring}, which characterizes the spread of many-body states in a selected basis of the Hilbert space and saturates at times $t^{(\mathrm{pe})}_{\mathrm{sat}} \propto \ln N$~\cite{Turkeshi24Delocalization}. The latter is tied to anticoncentration of the state of logarithmically deep random circuits~\cite{PRXQuantum.3.010333, bertoni2023shallow}, which is a necessary assumption of the formal proofs underlying quantum advantage. 
The rapid growth of non-stabilizerness is in a stark contrast with the ballistic increase of entanglement entropy under ergodic many-body dynamics~\cite{Kim13ballistic, nahum2017quantum}, resulting in a saturation timescale $t^{(\mathrm{ent})}_{\mathrm{sat}} \propto N$, linear in system size. At a formal level, the difference arises due to the disparity in boundary conditions at the top layer of the circuit corresponding to the GSEs~\eqref{eq:Ydtensor} and the entanglement entropy~\cite{zhou2019emergent} calculations. Physically, the time required to entangle two distant regions by local quantum dynamics scales linearly with the separation between the regions, consistent with the scaling of $t^{\mathrm{ent}}_{\mathrm{sat}}$. In contrast, the GSEs capture global properties of the state, and already time $t^{\mathrm{ent}}_{\mathrm{sat}} \propto \ln N$ is sufficient for nonstabilizerness to equilibrate even though entanglement between the most distant qudits has not yet been generated.

\section{Conclusions and outlook. }
In this paper, we have explored the dynamics of magic resources focusing on brick-wall Haar random unitary circuits. To that end, we considered the GSEs $M_W$ as scalable measures of nonstabilizerness, which include, but are not limited to, the stabilizer Rényi entropy. Our investigations reveal that magic resources are rapidly generated by the dynamics of random unitary circuits and saturate at relatively short times which scale logarithmically with the system size. The revealed behavior of $M_W$ aligns with the log-depth anticoncentration of random quantum circuits~\cite{PRXQuantum.3.010333} and matches the phenomenology of Hilbert space delocalization under random circuits~\cite{Turkeshi24Delocalization}. 
The GSE spreading remains qualitatively different from the ballistic growth of entanglement entropy in ergodic many-body systems. Since the random circuits constitute a minimal model of local unitary dynamics, we expect a similar phenomenology of magic state resources evolution to arise in generic ergodic many-body systems.

Understanding how the phenomenology of nonstabilizerness generation changes when the ergodicity is broken due to, e.g., many-body localization~\cite{abanin2019, sierant2024mbl} or quantum scars~\cite{Serbyn2021}, is an open question. Steps in that direction were already taken for integrable systems~\cite{rattacaso2023stabilizer,passarelli2024chaosmagicdissipativequantum, lópez2024exactsolutionlongrangestabilizer}, and doped Clifford circuits~\cite{haug2024probingquantumcomplexityuniversal} in which the generation of the magic resources is slower due to sparseness of  beyond-Clifford operations~\cite{supmat}. 
The asymmetry between the generation of magic resources and entanglement by local dynamics provides a new perspective onto the relation of entanglement and magic phase transitions~\cite{niroula2023phase,turkeshi2023errorresilience, bejan2023dynamical, fux2023entanglementmagic}. 
The framework based on the algebraic structure of the Clifford group, which yielded the GSEs, hosts more examples of magic measures with potential for better characterization of magic in many qudit systems. We leave these problems open for further research.

\section{Methods}
\label{sec:methods}

\subsection{Numerical simulations}
We employ two complementary numerical approaches: (i) exact circuit simulation and (ii) tensor network contraction for the annealed average of generalized stabilizer entropies. Together, they provide crucial insights into magic spreading in random quantum circuits. Exact simulation reveals the self-averaging properties of the GSEs, but is limited to small system sizes. Exploiting this fact, we use tensor network contraction to compute annealed averages, which accurately approximate quenched averages, as we argue in the following.

\subsubsection{Self-averaging}
\label{subsec:self}
Exact numerical simulation of quantum circuits' dynamics involves generating the computational basis $\mathcal B = \{ |\mathbf{x}_1 \mathbf{x}_2 \cdots \mathbf{x}_N\rangle \;|\; \mathbf{x}_{j}\in \mathbb{Z}_d\}$ and expressing the state $\ket{\Psi_t}$ as a superposition of the computational basis states. Then, the action of two-body Haar-random gates on states in  $\mathcal{H}_{N,d}$ 
reduces to sparse matrix-vector multiplication. 
The first limiting factor of the exact simulation is the exponential growth of $|\mathcal B | = d^N$ with the system size. 
A more severe constraint arises from the need to evaluate 
$| \mathcal P_N(d) |=d^{2N}$ Pauli string expectation values in the time-evolved state $\ket{\Psi_t}$ to compute the GSEs of interest, cf. Main Text. 
We developed and employed an efficient numerical algorithm~\cite{sieranttoappear}, allowing these exact calculations to system sizes up to  $N=22$ qubits and $N=12$ qutrits.

\begin{figure}
    \centering
    \includegraphics[width=1\linewidth]{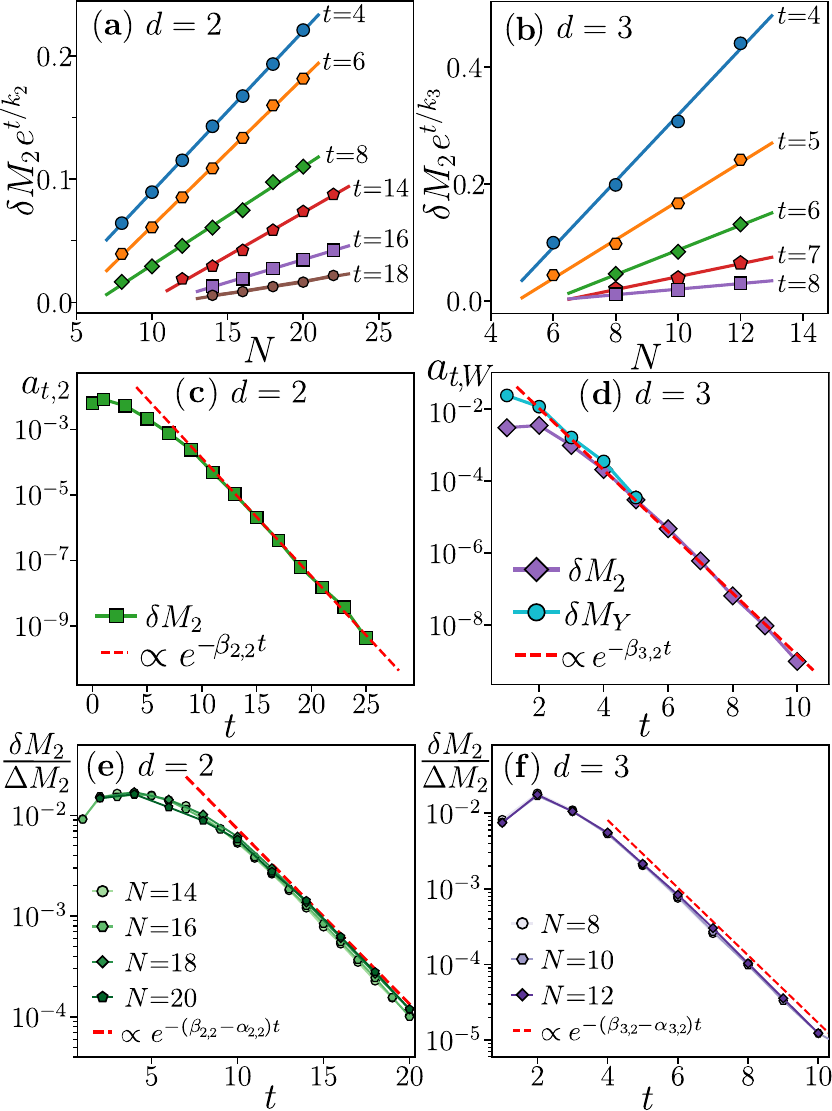}
    \caption{Self-averaging of SRE and GSE. 
    For qubits, (\textbf{a}),  and qutrits, (\textbf{b}), the difference $\delta M_2$ between the annealed and quenched averages of SRE scales linearly with the system size $N$, c.f., \eqref{eq:deltaM} (for presentation purposes $\delta M_2$ is rescaled by a factor $e^{t/k_d}$ with $k_2=3$ and $k_3=5/7$). The coefficient $a_t$, shown in (\textbf{c}) and (\textbf{d}), describing the leading term in this dependence, decays exponentially in time $t$, with characteristic rates $\beta_{2,2}=0.83(3)$ and $\beta_{3,2} = 1.97(5)$. The relative error decays exponentially in time, 
    \eqref{eq:delta2}, see (\textbf{e}) and (\textbf{f}).}
    \label{fig:self}
\end{figure}

We employ the exact numerical simulation to calculate the quenched average of the GSEs  $\overline{M}_W = -\mathbb{E}[\ln[\zeta_W(|\Psi_t\rangle)]]$ and the annealed average $\tilde{M}_W = -\ln[\mathbb{E}[\zeta_W(|\Psi_t\rangle)]]$, where the circuit average involves $1000$ realizations unless otherwise specified. 
The results are summarized in Fig.~\ref{fig:self}.

We first focus on the difference $\delta M_W(t) = |\overline M_W(t)-\tilde{M}_W(t)|$ between the quenched and annealed averages of GSEs. In Fig.~\ref{fig:self}(a),(b) we observe a clear linear system size dependence of the SRE  difference
\begin{equation}
\delta M_2(t) = a_{t,2} N + b_t,
\label{eq:deltaM}
\end{equation}
where $a_{t,2}$ and $b_{t,2}$ are constant at fixed time $t$. This behavior characterizes $\delta M_2(t)$ both for qubits ($d=2$) and qutrits ($d=3$). 
Analogous behavior is observed for $M_Y(t)$ (data not shown). 
As demonstrated in Fig.~\ref{fig:self}(c),(d), after an initial transient at small circuit depths $t$, the coefficients $a_{t,W}$ decreases exponentially over time for all relevant operators $ W $ in both qubit and qutrit systems
\begin{equation}
a_{t,W} = a_W e^{-\beta_{d,W} t},
\label{eq:at}
\end{equation}
where $\beta_{d,W}$ is a constant dependent on the on-site Hilbert space dimension $d$ and on $W$. 
For qubits, we find $\beta_{2,2} = 0.83(3)$ while for qutrits $\beta_{3,2} = \beta_{3,Y}= 1.97(5)$. These values of $\beta_{d,W}$ confirm that the error made when the quenched averages are interchanged with the annealed averages is negligible for sufficiently large $N$ and $t$. 
Indeed, the exponential decay of $\Delta M_W(t)$, as described in Eq.~\eqref{eq:decay}, occurs at a rate $\alpha_{2,2} = 0.43(3)$ for qubits and $\alpha_{3,2} = 1.05(3)$ for qutrits for $M_2$, while, for qutrits, $\Delta M_Y(t)$ decays with a rate of $\alpha_{3,Y} = 0.98(2)$.  

The rates $\beta_{d,W}$ are significantly (approximately twice) larger than the corresponding rates $\alpha_{d,W}$. Hence, the relative errors committed when the quenched and annealed averages are interchanged decays exponentially in time as 
\begin{equation}
    \delta M_W(t) /\Delta M_W(t) \propto e^{-(\beta_{d,W} - \alpha_{d,W})t},
    \label{eq:delta2}
\end{equation}
up to a sub-leading in system size term $O(1/N)$. These relative errors are shown in Fig.~\ref{fig:self}(e),(f) for the stabilizer R\'enyi entropy for $d=2,3$.  For any $t\geq 0$, the relative error is smaller than $3\%$ in all the considered cases. Moreover, we indeed observe a clear exponential decay of the relative error starting at times $\approx 5-10$ at which the saturation of GSEs is observed for the system sizes considered in this work. 

The above scaling analysis demonstrates that approximating quenched averages with the annealed averages in the dynamics of brick-wall quantum circuits is justified at any time  $t$ and that this approximation improves exponentially with $t$ as shown by~\eqref{eq:delta2}. Moreover, our numerical results indicate that the linear scaling~\eqref{eq:deltaM} is robust and persists at any system size $N$. These two trends show that our  main conclusion about the magic resources growth in random quantum circuits, i.e. the logarithm depth saturation of magic resources, $t^{\mathrm{mag}}_{\mathrm{sat}} \propto \ln(N)$, is accurate in the scaling limit of large 
system size $N$.

\begin{figure}
    \centering
    \includegraphics[width=1\linewidth]{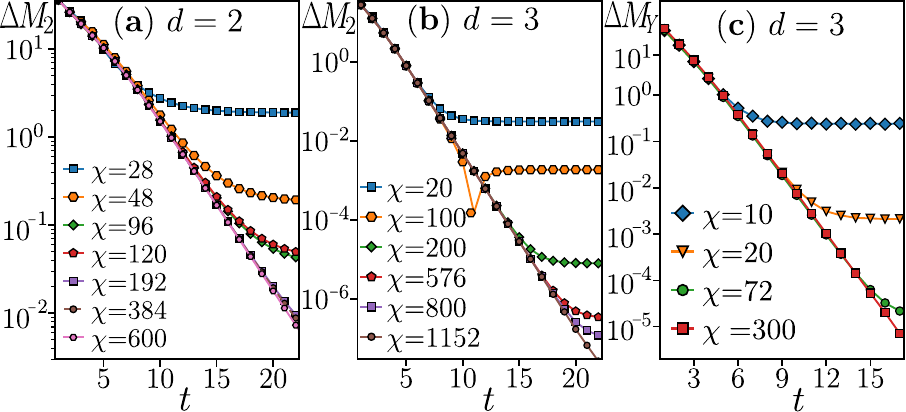}
    \caption{Convergence of the tensor network contraction for $\Delta{M}_2(t)$ and $\Delta M_Y(t)$ with bond dimension $\chi$ and systems comprising $N=64$ qubits (\textbf{a}) and qutrits $(\textbf{b},\, \textbf{c})$. 
    Increase of the bond dimension around $\chi = O(q_{\mathrm{eff}}^2)$ leads to no significant changes in the value of $\Delta M_W =M^{\mathrm{Haar}}_W -\tilde{M}_W(t)$, indicating the convergence of the results.}
    \label{fig:bond2}
\end{figure}

\subsubsection{Tensor network contractions}
\label{subsec:ten}

Our tensor network approach aims at an efficient evaluation of the tensor network contraction \eqref{eq:Ydtensor} which yields the annealed average 
$\tilde{M}_W(t)$. To that end, we interpret~\eqref{eq:Ydtensor} as a non-unitary time evolution of a state of $N$ effective "spins" with onsite Hilbert space dimension $q_{\mathrm{eff}}=k!$ growing factorially with the number $k$ of replicas. The non-unitary time evolution is followed by the contraction of the obtained state with the last layer~\eqref{eq:lastLayer}.

The state of the effective spins is expressed as a matrix product state (MPS) 
\begin{equation}
| \Pi \rrangle = \sum_{\tau_1,\dots,\tau_N \in S_k} A_{[1]}^{\tau_1} A_{[2]}^{\tau_2} \cdots A_{[N]}^{\tau_N} |\tau_1,\dots,\tau_N \rrangle,
\end{equation}
where $|\tau_1,\dots,\tau_N \rrangle = |\tau_1  \rrangle  \otimes |\tau_2  \rrangle \dots \otimes |\tau_N \rrangle$ is the product state of representations of permutations $\tau_i \in S_k$ while 
$A_{[i]}^{\tau_i}$ are $\chi \times \chi$ matrices for any $i=2, \dots, N-1$, $A_{[1]}^{\tau_1}$ ($A_{[N]}^{\tau_N}$) are $1 \times \chi$ ($\chi \times 1$) matrices and  $\chi$ is the bond dimension, which needs to be contrasted with the on-site local Hilbert dimension $q_{\mathrm{eff}}$. At $t=1$, state of the system $| \Pi_1 \rrangle$ is the product state \eqref{eq:initial}.

The time evolution consists of contraction of the state of the system $| \Pi_t \rrangle$ with subsequent layers of the tensors $\mathcal{T}^{(k)}_{i,i+1}$ defined by \eqref{eq:wten}. The contraction of $\mathcal{T}^{(k)}_{i,i+1}$ with the matrices $A_{[i],a_i,a_{i+1}}^{\tau_i}$, $ A_{[i+1],a_{i+1},a_{i+2}}^{\tau_{i+1} }$ (where $a_i$, $a_{i+1}$, $a_{i+2}$ denote the matrix indices of $A_{[i]},\, A_{[i+1]}$) results in a tensor $\mathcal{T}^{\tau_i,\tau_{i+1} }_{a_i,a_{i+2} }$. Reshaping the tensor to $\mathcal{T}_{(\tau_i, a_i),(\tau_{i+1}, a_{i+2} )}$ results in a matrix of dimension $(q_{\mathrm{eff}} \, \chi)\times (q_{\mathrm{eff}} \, \chi)$, which is expressed back as a product of two matrices $A'_{[i]},\, A'_{[i+1]}$ via the standard singular value decomposition (SVD). Notably, the dimension of the matrix $\mathcal{T}_{(\tau_i, a_i),(\tau_{i+1}, a_{i+2} )}$ is increased by a factor $q_{\mathrm{eff}}$ with respect to the bond dimension $\chi$. The complexity of SVD scales as $(q_{\mathrm{eff}} \, \chi)^3$ which is the main problem hindering our calculations for qubits ($d=2$), since the on-site Hilbert space dimension is $q_{\mathrm{eff}} = 4! =24$. This fact was one of our motivations for introducing the GSE $M_Y$ and considering qutrits, which results in a much smaller on-site Hilbert space dimension $q_{\mathrm{eff}}= 3! = 6$ and a significantly simpler computation of the tensor network contraction. Finally, for qubits ($d=2$), as noted in~\cite{Braccia2024}, the effective on-site Hilbert space dimension can be reduced from $q_{\mathrm{eff}}= k!$ to $q'_{\mathrm{eff}}= C_k$, where $C_k$ is the Catalan number. This enables us to calculate $\tilde{M}_2(t)$ for qubits by considering model with $q'_{\mathrm{eff}}=C_4=14$ dimensional on-site Hilbert space. To find the mapping between the tensors of \eqref{eq:Ydtensor} living $q_{\mathrm{eff}}= k!$ dimensional space to the $q'_{\mathrm{eff}}=14$ subspace, we employed the single qubit Haar-random gate which, upon averaging, corresponds to the projector onto the $14$-dimensional invariant subspace.

\begin{figure}
    \centering
    \includegraphics[width=1\linewidth]{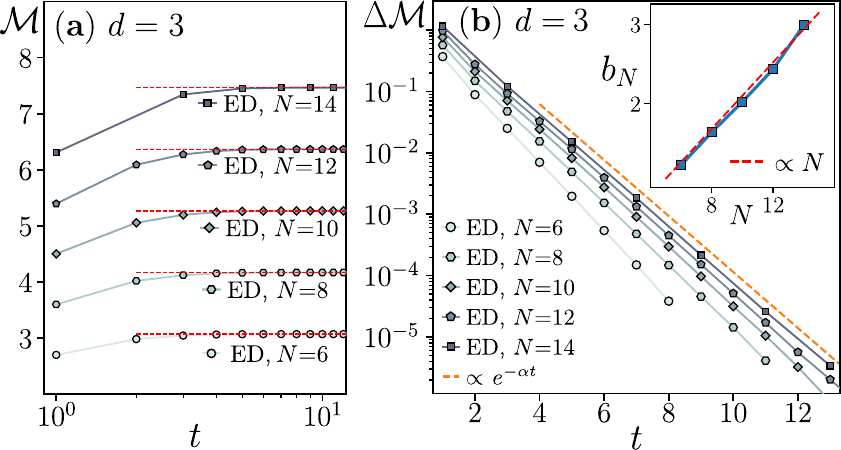}
    \caption{Dynamics of mana $\mathcal{M}$ for a system of $N$ qutrits ($d=3$) under Haar random circuits. (\textbf{a}) Saturation of $\mathcal{M}$ to $\mathcal{M}^{\mathrm{Haar}}$ occurs rapidly with the circuit depth $t$. (\textbf{b}) The difference $\Delta \mathcal{M} = \mathcal{M}^{\mathrm{Haar}} - \mathcal{M}$ follows $\Delta M_2 \propto N e^{-\alpha t}$ with $\alpha_3 = 1.05(9)$ at $t \gtrsim 4$. The annealed averages $\mathcal{M}$ over more than $100$ circuit realizations are calculated with the exact numerics (ED) for all $N$.}
    \label{fig:mana}
\end{figure}

In Fig.~\ref{fig:bond2}, we show the dependence of the tensor network contraction results for $\Delta \tilde{M}_W(t)$ for both qubits and qutrits as a function of the bond-dimension $\chi$. The results cease to be $\chi$-dependent once $\chi$ is of order of $q_{\mathrm{eff}}^2$.
Specifically, computation of the SRE $\Delta \tilde{M}_2(t)$ for qubits is well-converged with $\chi$ already at $\chi=192$, see Fig.~\ref{fig:bond2}(a). In contrast, convergence of the $\Delta \tilde{M}_2(t)$ calculation for times $t\in[0,18]$ for qutrits occurs only at $\chi=800$, as shown in Fig.~\ref{fig:bond2}(b). This demonstrates how the reduction $q_{\mathrm{eff}}=24 \to q'_{\mathrm{eff}}=14$, unavailable for qutrits, diminishes the computational resources in the calculations for qubits\footnote{Using irreducible representations reduces $q_\mathrm{eff} $ to $23$ for qutrits, maintaining similar limitations as the full $24$-dimensional case.}. 
Finally, Fig.~\ref{fig:bond2}(c) presents the calculations of $\Delta M_Y(t)$ for qutrits which require singificantly smaller resources and are converged for $t\in[0,15]$ already for $\chi=300$. 
By varying the system size, we have verified that the conclusions for the convergence with bond dimension are nearly independent of system size as long as $N \gtrapprox 30$. The results presented in the Main Text are converged with the bond dimension $\chi$.

\subsection{Results beyond generalized stabilizer entropies: mana}
\label{subsec:mana}

In the Main Text, we considered the growth of GSEs under dynamics of Haar random brick-wall circuits acting on systems of qubits and qutrits finding the exponential relaxation of these nonstabilizerness measures towards their long time values~\eqref{eq:decay}. Here, we argue that analogous phenomenology is shared by \textit{mana}, a nonstabilizerness monotone defined for odd prime $d$ as the negativity of the Wigner representation of the state $\rho$~\cite{Veitch2014theresourcetheory}.   

The phase-space point operators  $A_{\mathbf{r}}$ are defined in terms of the Pauli strings $P_{\bf{r}}= 
\prod_{l=1}^{N}X_l^{r_l^x} Z_l^{r_l^z}$ (where $\textbf{r} \equiv(r_1^x,r_1^z,\ldots,r_N^x,r_N^z)\in \mathbb{Z}_d^{2N}$) as 
\begin{equation} \label{eq:phase-space}
    A_{\mathbf{0}} = \frac{1}{d^N} \sum_\mathbf{r \in  \mathbb{Z}_d^{2N} } P_\mathbf{r}, \quad A_{\mathbf{r}} = P_{\mathbf{r}}  A_{\mathbf{0}} P_{\mathbf{r}}^\dagger.
\end{equation}
The phase-space point operators are orthogonal, $\mathrm{tr}(A_\mathbf{r} A_\mathbf{r'}) = d^N \delta_{\mathbf{r},\mathbf{r'}}$, enabling to represent the state as 
\begin{equation}
     \rho = \sum_\mathbf{r  \in  \mathbb{Z}_d^{2N} }  \frac{1}{d^N} \mathrm{tr}(A_{\mathbf{r}}\rho) A_{\mathbf{r}}.
\end{equation}
The coefficients of the expansion define the discrete Wigner function $W_{\rho}(\mathbf{r}) = \frac{1}{d^N} \mathrm{tr}(A_{\mathbf{r}}\rho)$~\cite{gross2006hudsonstheoremfor}.
Mana  $\mathcal{M}$ is defined as the negativity of the Wigner function
\begin{equation} \label{eq:mana}
    \mathcal{M} = \ln (\sum_{\mathbf{u}}| W_{\rho}(\mathbf{u})|).
\end{equation}
Mana offers insights into magic state resources as it is a strong nonstabilizerness monotone both for pure and mixed states. 

While calculating $\mathcal{M}$ does not require any costly minimization procedure, Eq.~\eqref{eq:mana} has no clear formulation within the replica formalism that would enable computing $\mathcal{M}$ in terms of a tensor network contraction~\eqref{eq:Ydtensor} for large systems comprising hundreds of qutrits. In the following, we resort to exact numerical computation of the time evolution of mana under brick-wall Haar random circuits for $N \leq 14$. Our results are shown in Fig.~\ref{fig:mana}. We observe that mana becomes extensive, $\mathcal{M} \propto N$ already at $t=1$, after a single layer of the circuit. Subsequently, $\mathcal{M}$ quickly saturates with $t$ to the value $\mathcal{M}^{\mathrm{Haar}}$ of the Haar-random state of $N$ qutrits. As shown in Fig.~\ref{fig:mana}~(b), the difference $\Delta \mathcal{M} = \mathcal{M}^{\mathrm{Haar}}-\mathcal{M}(t)$ exhibits a clear exponential decay, $\Delta \mathcal{M} = b_n e^{-\alpha t}$, with circuit depth $t$. 
The prefactor $b_n$ increases linearly with $N$, as shown in the inset in Fig.~\ref{fig:mana}~(b). Moreover, the numerical results suggest that the value of $\alpha$ is converging with the increase of $N$ towards a constant value. While the available range of $N$ is not sufficient for a clear demonstration that the saturation time of mana follows $t^{(\mathcal{M})}_{\mathrm{sat}} \propto \ln N$, the behavior of the coefficients $\alpha$ and $b_N$ strongly suggests that $\Delta \mathcal{M} \propto N e^{-\alpha t}$, analogously to the GSEs~\eqref{eq:decay}.  This demonstrates the universality of the uncovered phenomenology of magic spreading among different measures of nonstabilizerness.

\textbf{Acknowledgements.---}
We are grateful to D. Gross, L. Leone, and A. Hamma for enlightening discussions, R. Fazio for valuable collaborations and comments, and M. Lewenstein for collaborations on related topics. We thank the anonymous referee for their suggestion regarding irreducible representations. 
X.T. acknowledges DFG under Germany's Excellence Strategy – Cluster of Excellence Matter and Light for Quantum Computing (ML4Q) EXC 2004/1 – 390534769, and DFG Collaborative Research Center (CRC) 183 Project No. 277101999 - project B01. E.T. was supported by the MIUR Programme FARE (MEPH), by QUANTERA DYNAMITE PCI2022-132919, and by the EU-Flagship programme Pasquans2. E.T.  acknowledge the CINECA award under the ISCRA initiative, for the availability of high-performance computing resources and support. 
P.S. acknowledges support from the European Research Council AdG NOQIA; MCIN/AEI (PGC2018-0910.13039/501100011033, CEX2019-000910-S/10.13039/501100011033, Plan National FIDEUA PID2019-106901GB-I00, Plan National STAMEENA PID2022-139099NB, I00, project funded by MCIN/AEI/10.13039/501100011033 and by the “European Union NextGenerationEU/PRTR" (PRTR-C17.I1), FPI); QUANTERA MAQS PCI2019-111828-2); QUANTERA DYNAMITE PCI2022-132919, QuantERA II Programme co-funded by European Union’s Horizon 2020 program under Grant Agreement No 101017733); Ministry for Digital Transformation and of Civil Service of the Spanish Government through the QUANTUM ENIA project call - Quantum Spain project, and by the European Union through the Recovery, Transformation and Resilience Plan - NextGenerationEU within the framework of the Digital Spain 2026 Agenda; Fundació Cellex; Fundació Mir-Puig; Generalitat de Catalunya (European Social Fund FEDER and CERCA program, AGAUR Grant No. 2021 SGR 01452, QuantumCAT \ U16-011424, co-funded by ERDF Operational Program of Catalonia 2014-2020); the computing resources at Urederra and technical support provided by NASERTIC (RES-FI-2024-1-0043); Funded by the European Union. Views and opinions expressed are however those of the author(s) only and do not necessarily reflect those of the European Union, European Commission, European Climate, Infrastructure and Environment Executive Agency (CINEA), or any other granting authority. Neither the European Union nor any granting authority can be held responsible for them (EU Quantum Flagship PASQuanS2.1, 101113690, EU Horizon 2020 FET-OPEN OPTOlogic, Grant No 899794), EU Horizon Europe Program (This project has received funding from the European Union’s Horizon Europe research and innovation program under grant agreement No 101080086 NeQST); ICFO Internal “QuantumGaudi” project; European Union’s Horizon 2020 program under the Marie Sklodowska-Curie grant agreement No 847648; “La Caixa” Junior Leaders fellowships, La Caixa” Foundation (ID 100010434): CF/BQ/PR23/11980043.

\textbf{Data Availability.---}
The source code and the data will be publicly shared at publication, and linked to the printed version of the manuscript.

%\bibliography{filtered_bib}
\bibliographystyle{apsrev4-2}

\onecolumngrid
\newpage

\clearpage
\begin{center}
\textbf{\large Supplemental Material}
\end{center}

\setcounter{equation}{0}
\setcounter{figure}{0}
\setcounter{table}{0}
\setcounter{page}{1}
\setcounter{section}{0}
\renewcommand{\theequation}{S\arabic{equation}}
\setcounter{figure}{0}
\renewcommand{\thefigure}{S\arabic{figure}}
\renewcommand{\thepage}{S\arabic{page}}
\renewcommand{\thesection}{S\arabic{section}}
\renewcommand{\thetable}{S\arabic{table}}
\makeatletter

\renewcommand{\thesection}{\arabic{section}}
\renewcommand{\thesubsection}{\thesection.\arabic{subsection}}
\renewcommand{\thesubsubsection}{\thesubsection.\arabic{subsubsection}}

\vspace{0.3cm}

In this Supplemental Material, we discuss:
\begin{itemize}
    \item[(i)] properties of the Clifford group,
    \item[(ii)] properties of generalized stabilizer entropies (GSE),
    \item[(iii)] typical value of the GSEs of interest,
    \item[(iv)] efficient tensor network computability of the GSEs,
    \item[(v)] GSE growth under doped Clifford circuits.
\end{itemize}

\section{Selected properties of the Clifford group and its commutants}
The Clifford group, a subgroup of unitary operations $\mathcal{C}_{N,d} \subset \mathcal{U}(d^N)$, maps a Pauli string to another single Pauli string, up to a phase $C P C^\dagger = \omega^r P$, with $r \in \mathbb{Z}_d$.
It is completely generated by the Hadamard, CADD, and phase gates, respectively
\begin{equation}
\begin{split}
    H &= \frac{1}{\sqrt{d}} \sum_{m,n=0}^{d-1} \omega^{mn} |m\rangle\langle n|\;,\quad \mathrm{CADD} = \sum_{m,n=0}^{d-1} |m,m\oplus_d n\rangle\langle m,n|\;,\\
    P &= |0\rangle\langle 0| + i |1\rangle\langle 1|\ \text{ for $d=2$},\qquad P = \sum_{m=0}^{d-1} \omega^{m(m-1) 2^{-1}} |m\rangle\langle m|\ \text{ otherwise,}
\end{split}
\end{equation}
where $2^{-1}$ represents the multiplicative inverse in the finite field $\mathbb{Z}_d$.
With a slight abuse of notation, we will move between these notations, inferring the specific representation from the context. We define for later convenience $D=2d$ for $d=2$ and $D=d$ for $d\ge 3$.

The generalized stabilizer entropies (GSEs) are intimately connected to the commutant of the Clifford group, meaning those operators $W$ acting on $(\mathcal{H}_{d}^{\otimes N})^{\otimes k}$ such that $[W,C^{\otimes k}]=0$ for any $C\in\mathcal{C}_{N,d}$. We briefly review the key properties and results on the Clifford commutant, cf. Ref.~\cite{gross2021schurweylduality,montealegremora2022dualitytheoryfor} for an indepth analysis.
The Schur-Weyl duality links the $k$-commutant to the set $\Sigma_{k}(d)$ of stochastic Lagrangian subspaces of $\mathbb{Z}_d^{2k}$, namely the subspaces $T\subset \mathbb{Z}_d^{2k}$ such that: (i) any $(\mathbf{x},\mathbf{y})\in T$ has $\mathbf{x}\cdot \mathbf{x} - \mathbf{y}\cdot \mathbf{y} = 0 \;\mathrm{mod}\; D$, $T$ has dimension $\dim T = k$ in the field ${\mathbb{Z}^{2k}_d}$ and $\mathbf{1}_{2k}\equiv (1,1,\dots,1) \in T$.
These spaces induce operators $w(T) = \sum_{(\mathbf{x},\mathbf{y})\in T} |\mathbf{x}\rangle\langle \mathbf{y}|$ acting on $\mathcal{H}_d^{\otimes k}$, with $|\mathbf{x}\rangle = |x_1,\dots,x_k\rangle$ a vector associated with $\mathbf{x}\in \mathbb{Z}_d^k$. Similarly, we can construct an $N$-qudit operator in the $k$-replica space $W(T) = w(T)^{\otimes N}$. Both $w(T)$ and $W(T)$ are real matrices in the computational basis $\{ |\mathbf{x}\rangle \;|\; \mathbf{x}\in \mathbb{Z}_d^k\}$ and $\{ |\mathbf{x}^{(1)},\mathbf{x}^{(2)},\cdots,\mathbf{x}^{(N)}\rangle \;|\; \mathbf{x}^{(j)}\in \mathbb{Z}_d^k\}$, respectively.

To demonstrate that the operators $W(T)\in\mathrm{Comm}_k(\mathcal{C}_{N,d})$ for any $T\in \Sigma_k(d)$, it is sufficient to show that $[G, W(T)] = 0$ for the generators $G \in \{H, P, \mathrm{CADD}\}$, which span the whole Clifford group.
Consider the orthogonal space $T^\perp = \{(x,y) \in \mathbb{Z}_d^{2k} \; | \; \mathbf{x}\cdot \mathbf{x}' = \mathbf{y}\cdot \mathbf{y}' \;\mathrm{mod}\; d$ for any $(\mathbf{x},\mathbf{y}) \in T\}$. From the definition of stochastic Lagrangian subspaces, it follows that
\begin{equation}
    (\mathbf{x}+\mathbf{x}')\cdot (\mathbf{x}+\mathbf{x}') - (\mathbf{y}+\mathbf{y}')\cdot (\mathbf{y}+\mathbf{y}') = 0 \;\mathrm{mod}\; D = 2 (\mathbf{x}\cdot \mathbf{x}'-\mathbf{y}\cdot \mathbf{y}') \;\mathrm{mod}\; d = \mathbf{x}\cdot \mathbf{x}'-\mathbf{y}\cdot \mathbf{y}' \;\mathrm{mod}\; d\;,
\end{equation}
implying $T \subset T^\perp$, i.e., $T$ is self-orthogonal. Furthermore, since $\dim T + \dim T^\perp = 2k$ and $\dim T = k$, it follows that $\dim T^\perp = k$, meaning that $T = T^\perp$.
This remark allows us to show that, for the Hadamard gate
\begin{equation}
    H^{\otimes k} w(T) (H^\dagger)^{\otimes k} = \frac{1}{d^k} \sum_{\mathbf{m},\mathbf{n} \in \mathbb{Z}_d} \sum_{(\mathbf{x},\mathbf{y}) \in T} \omega^{\mathbf{m}\cdot \mathbf{x} - \mathbf{n}\cdot \mathbf{y}} |\mathbf{m}\rangle\langle \mathbf{n}| = \sum_{(\mathbf{x},\mathbf{y}) \in T^\perp} |\mathbf{x}\rangle\langle \mathbf{y}| = w(T).
\end{equation}
For the phase gate, simple algebraic manipulations in the qubit case where $d=2$
\begin{equation}
    P^{\otimes k} w(T) (P^\dagger)^{\otimes k} = \sum_{(\mathbf{x},\mathbf{y}) \in T} i^{\mathbf{x}\cdot \mathbf{x} - \mathbf{y}\cdot \mathbf{y}} |\mathbf{x}\rangle\langle \mathbf{y}| = \sum_{(\mathbf{x},\mathbf{y}) \in T} |\mathbf{x}\rangle\langle \mathbf{y}| = w(T),
\end{equation}
while for $d \ge 3$, a prime number, we have instead
\begin{equation}
    P^{\otimes k} w(T) (P^\dagger)^{\otimes k} = \sum_{(\mathbf{x},\mathbf{y}) \in T} \omega^{2^{-1} \sum_{m=0}^{d-1} [x_m(x_m-1) - y_m(y_m-1)]} |\mathbf{x}\rangle\langle \mathbf{y}| = \sum_{(\mathbf{x},\mathbf{y}) \in T} \omega^{-2^{-1}(\mathbf{x}\cdot \mathbf{1}_k-\mathbf{y}\cdot \mathbf{1}_k)} |\mathbf{x}\rangle\langle \mathbf{y}| = w(T).
\end{equation}
Finally, for the CADD, we consider $w(T)^{\otimes 2}$ and find
\begin{equation}
    \mathrm{CADD}^{\otimes k} w(T)^{\otimes 2} (\mathrm{CADD}^\dagger)^{\otimes k} = \sum_{(\mathbf{x},\mathbf{y}) \in T} \sum_{(\mathbf{x}',\mathbf{y}') \in T} |\mathbf{x}, \mathbf{x}+\mathbf{x}'\rangle \langle \mathbf{y}, \mathbf{y}+\mathbf{y}'| = \sum_{(\mathbf{x}, \mathbf{y}) \in T} \sum_{(\mathbf{z}, \mathbf{w}) \in T} |\mathbf{x}, \mathbf{z}\rangle \langle \mathbf{y}, \mathbf{w}| = w(T)^{\otimes 2},
\end{equation}
concluding the proof.

The operators $W(T)$ induced by the stochastic lagrangian subspaces completely characterize the commutant, since, for $N\ge k-1$, $|\Sigma_{k}(d)| = |\mathrm{Comm}_k({\mathcal{C}}_{N,d})| = \prod_{i=0}^{k-2} (d^i+1)$.
As a consequence, identifying the operators in $\mathrm{Comm}_k({\mathcal{C}}_{N,d})$ amounts to the characterization of $T \in \Sigma_{k}(d)$.
We note that $\Sigma_{k}(d)$ is not a group, and in particular, the operators $W(T)$ are not generally invertible.
We define the stochastic orthogonal group $\mathcal{O}_k$ as the set of $k \times k$ matrices with entries in the field $\mathbb{Z}$ such that $O \mathbf{x} \cdot O \mathbf{x} - \mathbf{x} \cdot \mathbf{x} = 0 \;\mathrm{mod}\; D$ for all $\mathbf{x} \in \mathbb{Z}_d^k$ and $O \mathbf{1}_k = \mathbf{1}_k \;\mathrm{mod}\; d$.
Any operator $O \in \mathcal{O}_k$ has its graph $T_O \equiv \{ (O\mathbf{x}, \mathbf{x}) \; | \; \mathbf{x} \in \mathrm{Z}_d^k\} \in \Sigma_{k}(d)$ and $w(O) \equiv w(T_O) $ is invertible and orthogonal.
From the group $\mathcal{O}_k$, we can define the left and right group actions $OT = \{ (O \mathbf{x}, \mathbf{y}) \; | \; (\mathbf{x}, \mathbf{y}) \in T\}$ and $TO = \{ (\mathbf{x}, O^T \mathbf{y}) \; | \; (\mathbf{x}, \mathbf{y}) \in T\}$, both elements in $\Sigma_{k}(d)$.
This action is consistent with the composition of the operators $W(T)$ and $W(O)$, namely $W(O) W(T) W(O') = W(OTO')$ for any $O, O' \in \mathcal{O}_k$ and $T \in \Sigma_{k}(d)$.
Therefore, the left and right actions allow for the decomposition of $\Sigma_{k}(d)$ into a disjoint union of double cosets
\begin{equation}
    \Sigma_{k}(d) = \mathcal{O}_k \sqcup \mathcal{O}_k T_1 \mathcal{O}_k \sqcup \cdots \sqcup \mathcal{O}_k T_r \mathcal{O}_k\;
\end{equation}
for some $r$, and $T_1, \dots, T_r \in \Sigma_{k}(d)$ as choices of representatives in different cosets, and $\mathcal{O}_k$ is regarded as a subset of $\Sigma_{k}(d)$ under the identification $O \mapsto T_O$. A relevant subgroup of $\mathcal{O}_k$ that is also stochastic and orthogonal are the permutation operators $\pi \in S_k\subset \mathcal{O}_k$, playing a fundamental role in the Schur-Weyl duality for the Haar group.
\textit{En passant}, we note that the intrinsic Clifford commutant contains all operators $W(T)$ induced by $T\in \Sigma_k(d)\setminus S_k$.

Remarkably, the number of subspaces $T_r$ that generate disjoing double cosets is bounded to $r\le k$~\cite{gross2021schurweylduality}.
To see this, we recall that all elements of $\Sigma_k(d)$ can be expressed in terms of defect subspaces.
Specifically, for any $T \in \Sigma_{k}(d)$, we define its left and right defect subspaces, respectively, $\mathcal{L}(T) \equiv \{\mathbf{x} \in \mathbb{Z}_d^k \; | \; (\mathbf{x}, \mathbf{0}) \in T\}$ and $\mathcal{R}(T) \equiv \{\mathbf{y} \in \mathbb{Z}_d^k \; | \; (\mathbf{0}, \mathbf{x}) \in T\}$. Furthermore, for every $\mathbf{y} \in \mathcal{R}(T)^\perp$, the map $\mathcal{J}[\mathbf{y}] = [\mathbf{x}(\mathbf{y})]$ such that $(\mathbf{x}(\mathbf{y}), \mathbf{y}) \in T$ is a well-defined defect isomorphism between $\mathcal{R}(T)$ and $\mathcal{L}(T)$.
A stochastic Lagrangian subspace $T$ is uniquely determined by the triplet $(\mathcal{L}(T), \mathcal{R}(T), \mathcal{J})$.
One can further show that $\mathcal{J}$ can be induced by a stochastic orthogonal matrix $O \in \mathcal{O}_k$. Since orthogonal transformations preserve rank, it follows that for any $T, T' \in \Sigma_{k,k}(d)$, we have $T' = OT$ for a certain $O \in \mathcal{O}_k T$ if and only if $\dim \mathcal{L}(T) = \dim \mathcal{L}(T')$ and $\mathbf{1}_k \in \dim \mathcal{L}(T)$ if and only if $\mathbf{1}_k \in \dim \mathcal{L}(T')$.
This remark implies that $\Sigma_{k}(d)$ can have at most $k$ cosets, as $\dim \mathcal{L}(T) \le k$ for any $T \in \Sigma_{k}(d)$.
The operators $w(T)$ induced by the stochastic Lagrangian subspace $T$ are expressible in terms of the coset states. For any defect subspace
\begin{equation}
    |\mathcal{A}, [\mathbf{x}]\rangle = \frac{1}{\sqrt{|\mathcal{A}|}} \sum_{\mathbf{z} \in \mathcal{A}} |\mathbf{x} + \mathbf{z}\rangle\;,
\end{equation}
which form an orthonormal family for $[\mathbf{x}] \in \mathfrak{A} = \mathcal{A}^\perp/\mathcal{A}$. Then for $T$ characterized by $(\mathcal{L}(T), \mathcal{R}(T), \mathcal{J})$, we have
\begin{equation}
    w(T) = |\mathcal{L}(T)| \sum_{[\mathbf{y}] \in \mathcal{R}(T)^\perp/\mathcal{R}(T)} |\mathcal{L}(T), [\mathbf{x}(\mathbf{y})]\rangle \langle \mathcal{R}(T), [\mathbf{y}]|\;.
\end{equation}

A convenient choice of operators are Calderbank-Shor-Steane (CSS) codes, characterized by the left and right defect subspaces coinciding, i.e., $\mathcal{A} \equiv \mathcal{L}(T) = \mathcal{R}(T)$. In this case, the projector onto the code space satisfies
\begin{equation}
    Q_{\mathcal{A}} = \frac{1}{|\mathcal{A}|^2} \sum_{\mathbf{p}, \mathbf{q} \in \mathcal{A}} Z_{\mathbf{q}} X_{\mathbf{p}} = \sum_{[\mathbf{x}] \in \mathfrak{A}} |\mathcal{A}, [\mathbf{x}]\rangle \langle \mathcal{A}, [\mathbf{x}]|\;,
\end{equation}
and, for the associated operator $w(T) = |\mathcal{A}| Q_{\mathcal{A}}$.
Any such stochastic Lagrangian subspace $T$ identifies a \emph{unique} double coset of $\Sigma_{k}(d)$.

Finally, the set of subspaces $\Sigma_{k}(d)$ can be endowed with a semigroup structure $\circ$ such that $T \mapsto w(T)$ is a representation
\begin{equation}
    w(T_1) w(T_2) = |\mathcal{A}_1 \cap \mathcal{A}_2| w(T_1 \circ T_2)\;.
\end{equation}
\emph{Case 1.---} When $T_1$ is associated with $O \mapsto T_1$, then $T_1 \circ T_2 \equiv OT$. A similar reasoning applies when $T_2$ is the subspace associated with an isometry.
\emph{Case 2.---} When $T_1$ and $T_2$ are associated with (coinciding left and right) defect subspaces $\mathcal{A}_1$ and $\mathcal{A}_2$ respectively, we define ${T}_1 \circ T_2$ as the stochastic Lagrangian subspace $T$ fixed by the triple
\begin{equation}
\begin{split}
    \mathcal{L}(T) &\equiv (\mathcal{A}_1 + \mathcal{A}_2) \cap \mathcal{A}_1^\perp,\\
    \mathcal{R}(T) &\equiv (\mathcal{A}_1 + \mathcal{A}_2) \cap \mathcal{A}_2^\perp,\\
    \mathcal{J} &\colon \mathcal{R}(T)^\perp / \mathcal{R}(T) \to \mathcal{L}(T)^\perp / \mathcal{L}(T), \quad [\mathbf{y}] \mapsto [\mathbf{x}(\mathbf{y})],
\end{split}
\end{equation}
where $\mathbf{x}$ is such that $\mathbf{x}-\mathbf{y} \in \mathcal{A}_1 + \mathcal{A}_2$.
Arbitrary elements $T_1$ and $T_2$ in $\Sigma_{k}(d)$ can be recast in the two cases above via an isometry $\mathcal{O}_k$.
In conclusion, the Clifford commutant is completely determined by the stochastic orthogonal group $\mathcal{O}_k$ and by at most $r\le k$ defect subspace projectors.
The above ingredient provides a complete characterization of the Clifford commutant.
In the Main Text, we inferred the dependence $W=W(T)$ of the stochastic lagrangian subspace $T$ to lighten the notation.
In Ref.~\cite{leonetoappear} a simplified but equivalent construction is presented.

\subsection{Identifying Defect Subspaces}
While the orthogonal group $\mathcal{O}_k$ on $d$ dimensional qudits are well known~\cite{gross2021schurweylduality}, the remaining terms in the Clifford commutant requires the computation of the defect subspaces corresponding to a CSS code.
The brute force cost of identifying such defect subspaces grows exponentially as $d^k$. This process involves examining integers $n = 0, \dots, d^k-1$, expressing each $n$ as $\sum_{i=1}^{k} b_i d^{i-1}$ in the $D$-basis, resulting in a vector $\mathbf{v}^{(n)} \equiv (b_1, b_2, \dots, b_k)$. Each vector is checked to satisfy the conditions $\mathbf{v}^{(n)} \cdot \mathbf{v}^{(n)} = 0 \;\mathrm{mod}\; D$ and $\mathbf{v}^{(n)} \cdot \mathbf{1}_k = 0 \;\mathrm{mod}\; d$.
We compile a list of such vectors $\mathcal{S} = \{\mathbf{v}^{(n_1)}, \mathbf{v}^{(n_2)}, \dots, \mathbf{v}^{(n_K)}\}$, reducing it to a number $\tilde{K}$ of linearly independent vectors.
Each subset $\mathcal{S}_\mathcal{A}$, for which any $\mathbf{x}, \mathbf{y} \in \mathcal{S}_\mathcal{A}$ satisfies (i) $\mathbf{x} \cdot \mathbf{y} = 0 \;\mathrm{mod}\; d$ and (ii) $\mathbf{z} = \mathbf{x} + \mathbf{y} \;\mathrm{mod}\; d$ with $\mathbf{z} \cdot \mathbf{z} = 0 \;\mathrm{mod}\; D$ and $\mathbf{z} \cdot \mathbf{1}_k = 0 \;\mathrm{mod}\; d$, defines a defect subspace $\mathcal{A} = \mathrm{span}(\mathcal{S}_\mathcal{A})$.

To illustrate this concept, consider the case $d=2$ and $k=5$. By brute force evaluation, we find:
\begin{equation}
    \tilde{\mathcal{S}} = \{ (1,1,1,1,0), (1,0,1,1,1), (1,1,0,1,1), (1,1,1,0,1), (1,1,1,1,0) \}.
\end{equation}
The only subset of mutually orthogonal vectors, representing defect subspaces, includes those with a single generator, such as:
\begin{equation}
    \mathcal{A}_{(1,1,1,1,0)} \equiv \mathrm{span}(\{(1,1,1,1,0)\}), \quad \mathcal{A}_{(1,1,1,0,1)} \equiv \mathrm{span}(\{(1,1,1,0,1)\}), \quad \mathrm{etc.}\;.
\end{equation}
A noteworthy case arises for $k=6$ replicas and $d=2$, where:
\begin{equation}
    \tilde{\mathcal{S}} = \{ (1,1,1,1,0,0), \;\text{+ permutations} \},
\end{equation}
resulting in $15$ elements. Each element of $\tilde{\mathcal{S}}$ generates a one-dimensional defect subspace, e.g., $\mathcal{A}_{(1,1,1,1,0,0)} \equiv \mathrm{span}(\{(1,1,1,1,0,0)\})$.
However, subsets $\mathcal{S}_\mathcal{A}$ with two elements induce $9$ distinct 2-dimensional defect subspaces, such as:
\begin{equation}
    \mathcal{A} = \mathrm{span}(\{ (1,1,1,1,0,0), (1,1,0,0,1,1) \}).
\end{equation}
Interestingly, $\mathbf{1}_6$ is not a defect subspace, but corresponds to the anti-identity matrix, an element of $\mathcal{O}_6$~\cite{gross2021schurweylduality}. Similarly, all transposition (swaps) correspond to vectors $(1,-1,0,0,\dots)$ and their permutations.
Similarly, for $d=3$ and $k=4, 5$, only one-dimensional defect subspaces exist. At $k=6$, however, we encounter a two-dimensional space generated by:
\begin{equation}
    \mathcal{A} = \mathrm{span}(\{ (1,1,1,0,0,0), (0,0,0,1,1,1) \}),
\end{equation}
and other permutations.
This notation translates to a recipe to write $W(T)$ in the Pauli basis in replica space, cf. also Ref.~\cite{leonetoappear} for an in-depth discussion. Any commutant operator amounts to
\begin{equation}
    W(T) = \sum_{P_1,\dots,P_r\in\mathcal{P}_N(d)} \frac{1}{d^{Nr}} P_1^{\otimes v_1}\otimes  P_2^{\otimes v_2}\otimes\cdots \otimes  P_r^{\otimes v_r} \otimes I^{\otimes k-\sum_{j}v_j},\label{eq:wdef}
\end{equation}
up to a permutation operators $W(T_\pi)$ with $\pi\in S_k\subset \Sigma_k(d)$. Here, the integers $v_j\in \mathbb{Z}_d$, determined by the stochastic Lagrangian subspace $ T $, are positive $ v_j > 0$, and satisfy  $\sum_j v_j \leq k$ and $\sum_{j} v_j = 0 \;\mathrm{mod}\; d$.

\section{Generalized stabilizer entropies as good measures of magic}
We consider a pure state $|\Psi\rangle$ and recall that, for any intrinsic Clifford commutant operator $W$ the definition of the GSEs and associated generalized stabilizer purities
\begin{equation}
    M_W(|\Psi\rangle) = -\log[\zeta_W(|\Psi\rangle)],\quad \zeta_W(|\Psi\rangle) = \mathrm{Tr}(W|\Psi\rangle\langle\Psi|^{\otimes k}).
\end{equation}
The GSEs posses the following properties as nonstabilizerness measures.

\noindent \textbf{Clifford invariance.---} Since $W\in \overline{\mathrm{Comm}}_k(\mathcal{C}_{N,d})\subset {\mathrm{Comm}}_k(\mathcal{C}_{N,d})$, for any $C\in \mathcal{C}_{N,d}$ a Clifford unitary
    \begin{eqnarray}\nonumber
        \zeta_W(C|\Psi\rangle) &=  \mathrm{tr}(W (C|\Psi\rangle\langle  \Psi|C^\dagger)^{\otimes k}) =\mathrm{tr}\left[(C^\dagger)^{\otimes k}W C^{\otimes k} (|\Psi\rangle\langle \Psi|)^{\otimes k} \right]  = \mathrm{tr}(W(|\Psi\rangle\langle \Psi|)^{\otimes k})  = \zeta_W(|\Psi\rangle)\;.
    \end{eqnarray}
Hence $\zeta_W$ and, thus, $M_W$ are invariant under Clifford conjugation.

 \noindent \textbf{Additivity.---}
    Consider two states $|\psi\rangle$ and $|\phi\rangle$ defined respectively in $\mathcal{H}_d^{\otimes N_A}$ and $\mathcal{H}_d^{\otimes N_B}$ with $N=N_A+N_B$. Since $W = w^{\otimes N}$ is a tensor product over qudits, we have
     \begin{equation}\begin{split}
         M_W(|\psi\rangle\otimes |\phi\rangle) &=  -\log\left[ w^{\otimes N} \left[(|\psi\rangle\otimes |\phi\rangle) (\psi|\otimes \langle \phi|)\right]^{\otimes k}\right]\;\\ \nonumber
          &=  -\log\left[ w^{\otimes N_A} (|\psi\rangle\langle \psi|)^{\otimes k}\right] -\log\left[ w^{\otimes N_B} (|\phi\rangle\langle \phi|)^{\otimes k}\right]  =M_W(|\psi\rangle)+ M_W(|\phi\rangle)
          \end{split}
     \end{equation}
where we used basic properties of the logarithm function.

\noindent \textbf{Faithfulness.---}
We note that $|\Psi\rangle\in \mathrm{STAB}_{N,d}$ iff $M_W(|\Psi\rangle)=0$ for $W\in \overline{\mathrm{Comm}}_k(\mathcal{C}_{N,d})$ is equivalent to $|\Psi\rangle\in \mathrm{STAB}_{N,d}$ iff $\zeta_W(|\Psi\rangle)=1$ for $W\in \overline{\mathrm{Comm}}_k(\mathcal{C}_{N,d})$.
Suppose that $|\Psi\rangle\in \mathrm{STAB}_{N,d}$ is a stabilizer state. This implies existence of a Clifford transformation such that $|0\rangle^{\otimes N} = C|\Psi\rangle$. Being $W$ in the commutant for a certain $T\in\Sigma_k(d)$, we have
\begin{equation}
    \zeta_W = \mathrm{tr}(W(T)|\Psi\rangle\langle\Psi|^{\otimes k}) = \langle 0|^{\otimes N k} W(T) |0\rangle^{\otimes N k}= \langle \mathbf{0}_k|w(T)|\mathbf{0}_k\rangle^{\otimes N}=1\;,
\end{equation}
where in the last step we used that $\mathbf{0}_k\in T$ for any $T\in \Sigma_k(d)$.
Vice-versa, consider $\zeta_W(|\Psi\rangle)=1$ for $W\in \overline{\mathrm{Comm}}_k(\mathcal{C}_{N,d})$. (For $W\in \mathrm{Comm}_k(\mathcal{U}(d^N))$  this is identically true for any state.)
Using permutation invariance $1=\zeta_W=\zeta_{\tilde{W}}$ with $\tilde{W}$ in the form Eq.~\eqref{eq:wdef}. Thus
\begin{equation}
    1=\zeta_W(|\Psi\rangle) = \mathrm{Tr}[\tilde{W}|\Psi\rangle\langle \Psi|^{\otimes k} ]= \frac{1}{d^{Nr}}\sum_{P_1,\dots,P_r} \prod_{j=1}^r\beta_j^{v_j}\;,\label{eq:dioc}
\end{equation}
where we used $|\Psi\rangle\langle \Psi| = \sum_{P} \beta_P P/d^N$ with $\beta_P = \langle \Psi|P^\dagger |\Psi\rangle$.
Furthermore, being a pure state, $\sum_{P} |\beta_P|^2=d^N$. This constraint, together with the absolute value of Eq.~\eqref{eq:dioc} implies that $|\beta_P|=1$ for $d^N$ terms and zero otherwise.
To fix the phases, we note that \( |\beta_P| = 0,1 \) in Eq.~\eqref{eq:dioc} induces an infinite hierarchy of constraints \( 1 = \zeta_{W^{(q)}}(|\Psi\rangle) \) when considering \( k \mapsto qk \) replicas, with
\begin{equation}
    W^{(q)}(T) = \sum_{P_1,\dots,P_r\in\mathcal{P}_N(d)} \frac{1}{d^{Nr}} P_1^{\otimes q v_1} \otimes P_2^{\otimes q v_2} \otimes \cdots \otimes P_r^{\otimes q v_r} \otimes I^{\otimes q k-\sum_{j} (q v_j)},\label{eq:wdef2}
\end{equation}
and all possible permutation in $S_{qk}$. Recalling \( P^d = I \), this hierarchy holds if and only if \( \beta_P = \omega^{m_P} \) for some \( m_P \in \mathbb{Z}_d \). Thus,
\(
|\Psi\rangle\langle\Psi| = \sum_{P} {\omega^{m_P} P}/{d^N},
\)
implying \( |\Psi\rangle \in \mathrm{STAB}_{N,d} \), as required.

\section{Monotones of magic state resource theory }
In the following, we analyze the properties of GSEs from the point of view of magic state resource theory. A real-valued function $f( \rho )$ is a stabilizer monotone if, for any state $\rho$ and a stabilizer protocol $\mathcal{E}: \rho \to \mathcal{E}(\rho)$
\begin{equation}
    f(\mathcal{E}(\rho)) \leq f(\rho).
    \label{eq:mono}
\end{equation}
Our manuscript focuses on the spreading of magic resources under many-body dynamics. Hence, we are mostly interested in the protocols $f(.)$ corresponding to unitary evolution. Nevertheless, to put our considerations into a broader context, we investigate the properties of GSEs under the general stabilizer protocols. Recently, we remark that, for qubits, SREs have been proved as monotone~\cite{leone2024stabilizerentropiesmonotonesmagicstate} when involving also the computational basis measurements. Here we generalize the proof to $d$ dimensional qudits, and reveal with concrete counterexamples that GSEs are not generally monotones.

\subsection{Stabilizer R\'enyi entropies are magic monotone for qudits}
In this section, we prove that the stabilizer R\'enyi entropy is a magic monotone for any qudit dimension $d$ prime. The idea is based on Ref.~\cite{leone2024stabilizerentropiesmonotonesmagicstate}, and required bounding the expectation value $P_\alpha(\Psi)\equiv d^{-N}\sum_{P\in \mathcal{P}_N} |\langle \Psi | P | \Psi\rangle|^{2\alpha}$ for a state on $N$ qudits $|\Psi\rangle = \sum_{i=0}^{d-1} \sqrt{p_i} |i\rangle\otimes |\phi_i\rangle$ with $\sum_{i=0}^{d-1} p_i = 1$, $\{|i\rangle\}_{i=0,\dots,d-1}$ on-site state in the computational basis, and $\{|\phi_i\rangle \}_{i=0,\dots,d-1}$ generic states on the remaining $N-1$ qudits.
In particular, we have that the stabilizer R\'enyi entropy is a magic monotone if the following Lemma holds:
\begin{theorem}
Consider $|\Psi\rangle=\sum_{i=0}^{d-1} \sqrt{p_i} |i\rangle\otimes |\phi_i\rangle$ and $|\phi_i\rangle\in\mathbb{C}^{d\otimes (N-1)}$. For any integer $\alpha\ge 2$ it holds that
    \begin{equation}
    P_\alpha(\Psi)\le \max_{j}\{P_\alpha(\phi_j)\}\;.
\end{equation}
\end{theorem}

\begin{proof}
    Expanding $P_\alpha(\Psi)$ over $\tilde{P}\otimes P$ with $P\in \mathcal{P}_{N-1}$ and $\tilde{P}\in \mathcal{P}_1 $ we have
    \begin{align}
        P_\alpha(\Psi) & = \frac{1}{d^N} \sum_{P\in \mathcal{P}_{N-1} }\sum_{\tilde{P}\in \mathcal{P}_1} \Bigg|\sum_{i=0}^{d-1} \sum_{j=0}^{d-1} \langle i|\tilde{P}|j\rangle\langle \phi_i|P|\phi_j\rangle\sqrt{p_i p_j}\Bigg|^{2\alpha} \\
        & = \frac{1}{d^N} \sum_{P\in \mathcal{P}_{N-1} }\sum_{a,b=0}^{d-1} \Bigg|\sum_{i=0}^{d-1} \sum_{j=0}^{d-1} \langle i|X^a Z^b|j\rangle\langle \phi_i|P|\phi_j\rangle\sqrt{p_i p_j}\Bigg|^{2\alpha} \\
        & = \frac{1}{d^N} \sum_{P\in \mathcal{P}_{N-1} }\sum_{a,b=0}^{d-1} \Bigg|\sum_{i=0}^{d-1} \sum_{j=0}^{d-1} \langle i|j\oplus a \rangle \omega^{b j}\langle \phi_i|P|\phi_j\rangle\sqrt{p_i p_j}\Bigg|^{2\alpha} \\
        & = \frac{1}{d^N} \sum_{P\in \mathcal{P}_{N-1} }\sum_{a,b=0}^{d-1} \Bigg|\sum_{j=0}^{d-1}  \omega^{b j}\langle \phi_i|P|\phi_j\rangle\sqrt{p_{j\oplus a} p_j}\Bigg|^{2\alpha} \\
        & = \frac{1}{d^N} \sum_{P\in \mathcal{P}_{N-1} }\sum_{a,b=0}^{d-1}
        \sum_{|\vec{i}|=\alpha,|\vec{j}|=\alpha} \binom{\alpha}{\vec{i}}\binom{\alpha}{\vec{j}} \left(\prod_{m=0}^{d-1} \omega^{b m (j_m - i_m)}\right) \left(\prod_{m=0}^{d-1} \langle \phi_{m\oplus a}|P|\phi_m\rangle^{j_m} \langle \phi_{m}|P^\dagger|\phi_{m\oplus a}\rangle^{i_m} \right) \left(\prod_{m=0}^{d-1} (p_{j\oplus a} p_j)^{j_m+i_m} \right)\;,\nonumber
    \end{align}

where in the second step we substituted $\tilde{P}=X^a Z^b$ for $a,b=0,\dots,d-1$, and in the last step we introduced the multinomial indices $\vec{i}=(i_0,\dots,i_{d-1})$ with $0\le i_m\le 1$ and $|\vec{i}|\equiv \sum_{m=0}^{d-1}j_m$, while
\begin{equation}
    \binom{\alpha}{\vec{i}} = \frac{\alpha!}{\prod_{m=0}^{d-1} i_m!}\;.
\end{equation}

Summing over the $b$, we have a delta function forcing $\sum_{m=0}^{d} m (j_m-i_m) = 0 \mod d$. We denote this condition $\tilde\delta_{\vec{i},\vec{j}} \equiv \delta_{\sum_m m (j_m-i_m) = 0\mod d} $ for convenience. At the same time, we define $f^{(a)}_{\vec{i},\vec{j}}(\vec{p})\equiv \prod_{m=0}^{d-1} (p_{j\oplus a} p_j)^{j_m+i_m}$ for later convenience. Collecting these results, we have
\begin{equation}
    P_\alpha(\Psi) =  \frac{1}{d^{N-1}} \sum_{P\in \mathcal{P}_{N-1} }  \sum_{a=0}^{d-1}
        \sum_{|\vec{i}|=\alpha} \sum_{|\vec{j}|=\alpha} \binom{\alpha}{\vec{i}}\binom{\alpha}{\vec{j}} \tilde{\delta}_{\vec{i},\vec{j}}\left(\prod_{m=0}^{d-1} \langle \phi_{m\oplus a}|P|\phi_m\rangle^{j_m} \langle \phi_{m}|P^\dagger|\phi_{m\oplus a}\rangle^{i_m} \right) f^{(a)}_{\vec{i},\vec{j}}(\vec{p})\;.
\end{equation}
Until now all computations are exact. We now bound the expression taking the modulus. We have
\begin{align}
    \Bigg|\sum_{P\in\mathcal{P}_{N-1}}\prod_{m=0}^{d-1} \langle \phi_{m\oplus a}|P|\phi_m\rangle^{j_m} \langle \phi_{m}|P^\dagger|\phi_{m\oplus a}\rangle^{i_m} \Bigg|&\le \sum_{P\in\mathcal{P}_{N-1}} \prod_{m=0}^{d-1}\Bigg| \langle \phi_{m\oplus a}|P|\phi_m\rangle^{j_m} \langle \phi_{m}|P^\dagger|\phi_{m\oplus a}\rangle^{i_m} \Bigg|\\
    & \le \sum_{P\in\mathcal{P}_{N-1}} \prod_{m=0}^{d-1} \Big| \langle \phi_{m\oplus a}|P|\phi_m  \rangle\Big|^{j_m+i_m} \\
    & \le \prod_{m=0}^{d-1}\left(\sum_{P\in \mathcal{P}_{N-1}} |\langle \phi_{m} | P |\phi_{m\oplus a}\rangle|^{2\alpha}\right)^{\frac{i_m+j_m}{2\alpha}}\;,\label{eq:finn2}
\end{align}
where in the last step we used Holder's inequality.
Next, by manipulating the Pauli strings we get
\begin{equation}
    \sum_{P\in \mathcal{P}_{N-1}} |\langle \phi_{m} | P |\phi_{m\oplus a}\rangle|^{2\alpha}\le \left(\sum_{P\in \mathcal{P}_{N-1}} |\langle \phi_{m} | P |\phi_{m}\rangle|^{2\alpha}\right)^{1/2} \left(\sum_{P\in \mathcal{P}_{N-1}} |\langle \phi_{m\oplus a} | P |\phi_{m\oplus a}\rangle|^{2\alpha}\right)^{1/2}.\label{eq:finn}
\end{equation}
This follows expanding the expression
\begin{align}
    \sum_{P\in \mathcal{P}_{N-1}} |\langle \phi_{m} | P |\phi_{m\oplus a}\rangle|^{2\alpha} &= \sum_{P\in \mathcal{P}_{N-1}}  \mathrm{tr}(P |\phi_m\rangle\langle \phi_m| P^\dagger |\phi_{m\oplus a}\rangle\langle \phi_{m\oplus a}|)\\
    &= \frac{1}{d^{\alpha (N-1)}} \sum_{P} \sum_{P_1,\dots,P_\alpha}\prod_{i=1}^\alpha \left(\mathrm{tr}(P_i |\phi_m\rangle\langle\phi_m|)\mathrm{tr}(P^\dagger_i |\phi_{m\oplus a}\rangle\langle\phi_{m\oplus a}|)\right)\prod_{i=1}^\alpha K(P,P_i)\;,
\end{align}
where $K(P,Q) = \mathrm{tr}(P QP^\dagger Q^\dagger)/d^{N-1}$. Since $K(P,Q_1)K(P,Q_2)=K(P,Q_1 Q_2)$ we have
\begin{equation}
    \sum_{P} \prod_{i=1}^\alpha K(P,P_i) = \sum_{P} K(P,P_1\dots P_\alpha) = d^{2(N-1)} \delta_{P_1\dots P_\alpha,\mathbb{I}}.
\end{equation}
Thus
\begin{equation}
     \sum_{P\in \mathcal{P}_{N-1}} |\langle \phi_{m} | P |\phi_{m\oplus a}\rangle|^{2\alpha} = \frac{1}{d^{(\alpha-2)(N-1)}} \sum_{P_1,\dots,P_\alpha} \delta_{P_1\dots P_\alpha,\mathbb{I}}\prod_{i=1}^\alpha \langle \phi_m |P_i|\phi_m\rangle \langle \phi_m |P^\dagger_i|\phi_m\rangle\;,
\end{equation}
obtaining Eq.~\eqref{eq:finn} after using Cauchy-Schwarz for the $d^{2\alpha(N-1)}$ dimensional vectors with components
\begin{equation}
    \delta_{P_1\dots P_\alpha,\mathbb{I}} \mathrm{tr}(P_1|\phi_k\rangle\langle \phi_k|)\dots \mathrm{tr}(P_\alpha|\phi_k\rangle\langle \phi_k|)\;.
\end{equation}

From Eq.~\eqref{eq:finn2}, multiplying and dividing by factors $d^{N-1}$, recognizing the definition of $P_\alpha$ and collecting these statements we obtain
\begin{equation}
    P_\alpha(\Psi) \le \sum_{a=0}^{d-1}
        \sum_{|\vec{i}|=\alpha} \sum_{|\vec{j}|=\alpha} \binom{\alpha}{\vec{i}}\binom{\alpha}{\vec{j}} \tilde{\delta}_{\vec{i},\vec{j}}\prod_{m=0}^{d-1} \left( P_\alpha(\phi_m)^{\frac{i_m+j_m}{4\alpha}}P_\alpha(\phi_{m\oplus a})^{\frac{i_m+j_m}{4\alpha}} \right) f^{(a)}_{\vec{i},\vec{j}}(\vec{p})\;.
\end{equation}
Denoting $\max_{m} \{P_\alpha(\phi_m)\}$ the maximal value of $P_\alpha$ over the $|\phi_m\rangle$ states, we have
\begin{equation}
    P_\alpha(\Psi) \le \max_{m} \{P_\alpha(\phi_m)\} \sum_{|\vec{i}|=\alpha} \sum_{|\vec{j}|=\alpha} \binom{\alpha}{\vec{i}}\binom{\alpha}{\vec{j}} \tilde{\delta}_{\vec{i},\vec{j}} \sum_{a=0}^{d-1}f^{(a)}_{\vec{i},\vec{j}}(\vec{p})\;.
\end{equation}
We note that
\begin{equation}
    \sum_{a=0}^{d-1}f^{(a)}_{\vec{i},\vec{j}}(\vec{p}) \le d^{1-2\alpha}\;,
\end{equation}
and furthermore
\begin{equation}
    \sum_{|\vec{i}|=\alpha} \sum_{|\vec{j}|=\alpha} \binom{\alpha}{\vec{i}}\binom{\alpha}{\vec{j}} \tilde{\delta}_{\vec{i},\vec{j}} d^{1-2\alpha}\leq 1.
\end{equation}
Summarizing these statements, we have
\begin{equation}
    P_\alpha(\Psi) \le \max_{m} \{P_\alpha(\phi_m)\} \sum_{|\vec{i}|=\alpha} \sum_{|\vec{j}|=\alpha} \binom{\alpha}{\vec{i}}\binom{\alpha}{\vec{j}} \tilde{\delta}_{\vec{i},\vec{j}} \sum_{a=0}^{d-1}f^{(a)}_{\vec{i},\vec{j}}(\vec{p})\leq \max_{m} \{P_\alpha(\phi_m)\}
\end{equation}
concluding the proof of the Lemma. The monotonicity of the stabilizer R\'enyi entropy follows as a corollary.
\end{proof}

\subsection{Generalized Stabilizer entropies are not monotones}
To study the monotonicity of GSE under computational basis measurements, we consider (without loss of generality) a measurement of the first qutrit and expand the pure state $\ket{\Psi}$ of $N$ qudits as
\begin{equation}
  |\Psi\rangle = \sum_{i=0}^{d-1} \sqrt{p_i} |i\rangle\otimes |\phi_i\rangle,
  \label{eq:qudit}
\end{equation}
 where $\{\ket{i}\}_{i=0}^{d-1}$ are the computational basis states, $\sum_{i=0}^{d-1} p_i =1$, and  $\ket{\phi_i}$ are the states of the remaining $N-1$ qudits.

In the following, by analyzing the behavior of $M_Y$ under measurements in the computational basis, we demonstrate that this GSE \textit{is not} a monotone. To that end we find an example of a $N=2$ qutrits state for which \eqref{eq:mono} is not satisfied and $M_Y$ increases when a measurement in the computational basis is performed. We consider a state $\ket{\Psi_2} = \sum_{i,j=0}^{2} a_{i,j}\ket{i}\otimes\ket{j}$, where
$a_{00} = 0.04899 + 0.29503i$,
$a_{10} = 0.11566 + 0.04942i$,
$a_{20} = 0.24715 + 0.34531i$,
$a_{01} = 0.37355 + 0.23511i$,
$a_{11} = 0.15912 + 0.33928i$,
$a_{21} = 0.33144 + 0.08953i$,
$a_{02} = 0.10273 + 0.08267i$,
$a_{12} = 0.43260 + 0.22725i$,
$a_{22} = 0.02948 + 0.06498i$). For this state, we have $Y_3( \ket{\Psi_2}\bra{\Psi_2} )\approx 0.7837$, which is \textit{smaller} than the GSE values for each of the states obtained after the measurement of the first qutrits in computational basis: $Y_3( \ket{\phi_1}\bra{\phi_1} )  \approx 1.0004$, $Y_3( \ket{\phi_2}\bra{\phi_2} )  \approx 1.0408$, $Y_3( \ket{\phi_3}\bra{\phi_3} )  \approx 1.4456$. This counter-example shows that not all nonstabilizerness measures induced by the defect subspaces of the Clifford commutant are monotones.

\section{Generalized stabilizer entropy of Haar Random States}
We discuss the GSEs of Haar random states. Due to the concentration of the Haar measure on the unitary group for $N \gg 1$, we have $\overline{M_W} \equiv \mathbb{E}_\mathrm{Haar}[\zeta_W] = -\log [\mathbb{E}_\mathrm{Haar}[\zeta_W]] + O(\exp(-\gamma N))$.
Recall from Ref.~\cite{gross2021schurweylduality} that:
\begin{equation}
   \mathbb{E}_\mathrm{Haar}[|\Psi\rangle\langle \Psi|^{\otimes k}] = \frac{(d^N-1)!}{(d^N+k-1)!}\sum_{\pi \in S_k} W(T_\pi),
\end{equation}
where $T_\pi$ is the stochastic Lagrangian subspace induced by the permutation $\pi \in S_k \subset \mathcal{O}_k$.
Thus, we find for the $W=W(T)$ of interest
\begin{equation}
    \overline{\zeta_{W(T)}} = \mathbb{E}_\mathrm{Haar}[\zeta_{W(T)}] =\frac{(d^N-1)!}{(d^N+k-1)!} \sum_{\pi \in S_k} \left[\mathrm{tr}(w(T) w(T_\pi))\right]^N=\frac{(d^N-1)!}{(d^N+k-1)!} \sum_{\pi \in S_k} d^{N|T\cap T_\pi|}\;,
\end{equation}
where $|T\cap T_\pi|$ is the cardinality of the intersection between the stochastic lagrangian subspaces.
We specialize to the case of \( M_2 \) for qubits and qutrits with \( k=4 \) replicas, and \( M_Y \) for qutrits with \( k=3 \) replicas.
In these cases we have
\begin{equation}
    \mathrm{tr}(w(T) w(T_\pi))
    = \begin{cases}
        d^{D-1} & \pi = (),\\
        d^{2} & \pi \in \{(12 \dots D),(23\dots D1),\dots,(D1\dots 2)\},\\
        d^{3} & \pi \in S_2 \otimes S_2, \text{ and } d=2,\\
        d^{3} & \pi \in \{(13)(24),(14)(23)\} \text{ and } d=3,\\
        d & \pi =(12)(34)\text{ and } d=3,\\
        d^{\#(\pi)-1} & \text{otherwise,}
    \end{cases}
\end{equation}
where $\#(\pi)$ is the number of cycles. Combining these results, we recast $M_W^\mathrm{Haar}$ for the GSEs of interest in the Main Text.

\section{Tensor Network Computability}
In this section, we argue that GSEs can be efficiently computed for any defect subspace $\mathcal{A}$ using tensor network protocols.
A matrix product state (MPS) $|\Psi_N\rangle$ is defined by
\begin{equation}
|\Psi_N\rangle = \sum_{\{s_k \in \mathbb{Z}_d\}_k} A_{[1]}^{s_1} A_{[2]}^{s_2} \cdots A_{[N]}^{s_N} |s_1, \dots, s_N\rangle,
\end{equation}
where $A_i^{s_i}$ are $\chi_i \times \chi_{i+1}$ matrices for any $i=2, \dots, N-1$, and $A_1^{s_1}$ ($A_N^{s_N}$) are $1 \times \chi_2$ ($\chi_N \times 1$) matrices. Here, in contrast to the Main Text, we are considering the state of $N$ qudits. For simplicity, we assume $\chi = \chi_i$ for any $i = 1, \dots, N$, and refer to $\chi$ as the bond dimension.
The following ideas extend straightforwardly to translationally-invariant states.
As discussed in the Main Text, $\zeta_W = \mathrm{tr}(w^{\otimes N} |\Psi\rangle\langle \Psi|^{\otimes k})$. Given that $w \ge 0$, it can be decomposed via $w = \Gamma^\dagger \Gamma$.
This decomposition allows us to define a new tensor
\begin{equation}
    B_{[i]}^{\tilde{s}_i} = \Gamma \cdot A_i^{\otimes k},
\end{equation}
with a bond dimension $\chi^k$ and a physical dimension $\mathrm{rank}(w)$. The corresponding replica MPS is
\begin{equation}
    |\Phi^{(N)}\rangle \equiv \sum_{\{\tilde{s}_k \in \mathcal{A}\}_k} B_{[1]}^{\tilde{s}_1} B_{[2]}^{\tilde{s}_2} \cdots B_{[N]}^{\tilde{s}_N} |\tilde{s}_1, \dots, \tilde{s}_N\rangle,
\end{equation}
and the computation of magic resources simplifies to computing the norm
\begin{equation}
    M_W(|\Psi_N\rangle) = -\log\langle \Phi^{(N)}| \Phi^{(N)}\rangle.
\end{equation}
In a similar fashion, the MPS formulation facilitates efficient sampling methods, such as Pauli Monte-Carlo and perfect sampling, as well as compression algorithms for tensor networks.

\section{Additional Numerical results}
\subsection{Stabilizer R\'enyi entropy growth under doped Clifford circuits}

In the Main Text, we argued that the brick-wall random Haar circuits serve as minimal models of generic many-body dynamics. In particular, each layer of the Haar circuit contains extensively many 2-body gates, each generating a non-vanishing amount of the GSE.

There are, however, instances in which the resources generating non-stabilizerness are much more sparse which hinders the growth of the quantum magic resources. To illustrate this point, we consider an example of a doped Clifford circuit, in which 2-qubit unitary gates $U^{(C)}_{i,i+1}$ that form a brick-wall lattice are drawn with uniform probability from the Clifford group $\mathcal{C}_{2,2}$. The consecutive layers of the circuit are interspersed with action of a randomly placed T-gate defined as $T\ket{m} = e^{-i \pi m/4}\ket{m}$, which is the sole non-stabilizerness generating ingredient of the circuit's dynamics.

\begin{figure}[ht]
    \centering
    \includegraphics[width=0.8\columnwidth]{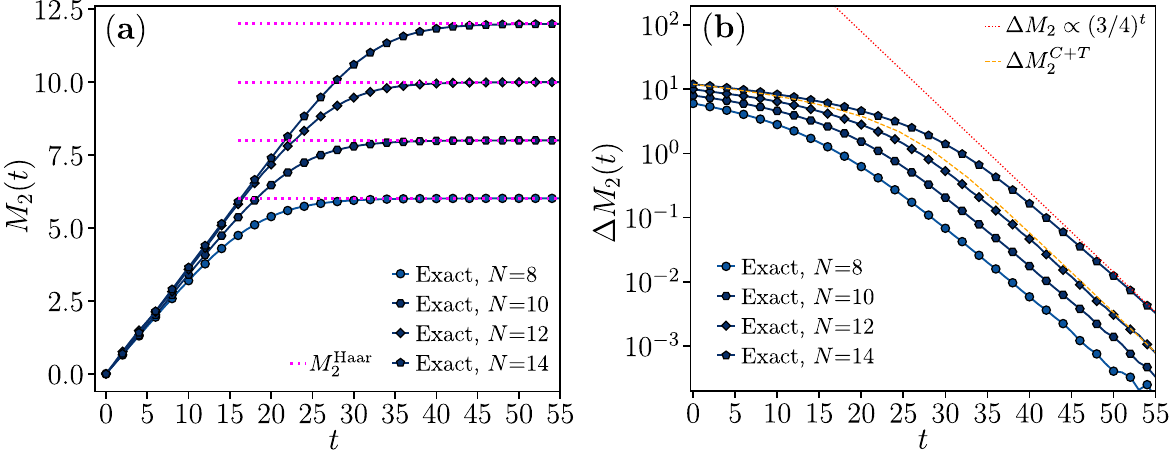}
    \caption{Doped Clifford circuit simulations. (a) The evolution of $Y_2(t)$ obtained via exact numerics for system sizes $N\le 14$. The magenta dotted lines are the value at $t\to \infty$. (b) Crossover behavior of the difference $\Delta M_2(t) \equiv M^{\mathrm{Haar}}_2 - M_2(t)$ to a regime of exponential decay, $\Delta M_2 \propto (3/4)^t$ (denoted by the red dotted line).
    The formula $\Delta M^{C+T}_2$ \eqref{eq:Y2ct} (denoted by the orange dashed line) qualitatively reproduces the bahavior of $\Delta M_2$. The results are averaged over more than $1000$ circuit realizations.
    }
    \label{fig:CpT}
\end{figure}

Our exact numerical results for the evolution of the SRE $M_2(t)$ obtained for this setup are summarized in Fig.~\ref{fig:CpT}. At small times, the SRE increases linearly with time, reflecting the gradual generation of magic resources by the $T$-gate and their spreading due to the Clifford dynamics. At long times, the SRE saturates to value $M^{\mathrm{Haar}}_2$, and the approach to the saturation value, captured by $\Delta Y_2(t) = M^{\mathrm{Haar}}_2- M_2(t)$, becomes exponential in time, $\Delta M_2(t) \propto e^{-\alpha_C t} $, where $\alpha_C>0$ is a constant. Incidentally, we observe that our numerical results for doped Clifford circuits are qualitatively reproduced by the formula
\begin{equation}
    \Delta M^{C+T}_2(t) = M^{\mathrm{Haar}}_2 + \log[M^{\mathrm{Haar}}_2 +(3/4)^t]
    \label{eq:Y2ct},
\end{equation}
obtained in \cite{leone2022stabilizerrenyientropy, haug2024probingquantumcomplexityuniversal} for circuits in which each layer of the Clifford gates is replaced by a global, $N$-qubit, Clifford gate. While Eq.~\eqref{eq:Y2ct} corresponds to a circuit in which the non-stabilizerness spreading is more efficient, the linear growth of $M_2(t)$ at small times, as well as, the crossover to a regime of exponential relaxation towards the saturation value  $M^{\mathrm{Haar}}_2$ are reflected by this equation. The latter behavior resembles the phenomenology of the brick-wall Haar random quantum circuits considered in the Main Text. Notably, however, the saturation of the SRE up to a fixed tolerance occurs at $t^{(C+T)}_{\mathrm{sat}} \propto N$, i.e., at times much longer than $t^{\mathrm{mag}}_{\mathrm{sat}}$ scaling logarithmically with $N$ for generic many-body dynamics.

The doped Clifford circuits are, however, fine-tuned, i.e., a generic small perturbation $\delta U$ of the Clifford gate $U^{(C)}_{i,i+1}$ results in a gate $U^{(C)}_{i,i+1}+\delta U$ which is no longer Clifford. Such a perturbed system is expected to follow the phenomenology of Haar circuits discussed in the Main Text.

\subsection{R\'{e}nyi index dependence}
\begin{figure}[ht]
    \centering
    \includegraphics[width=0.8\columnwidth]{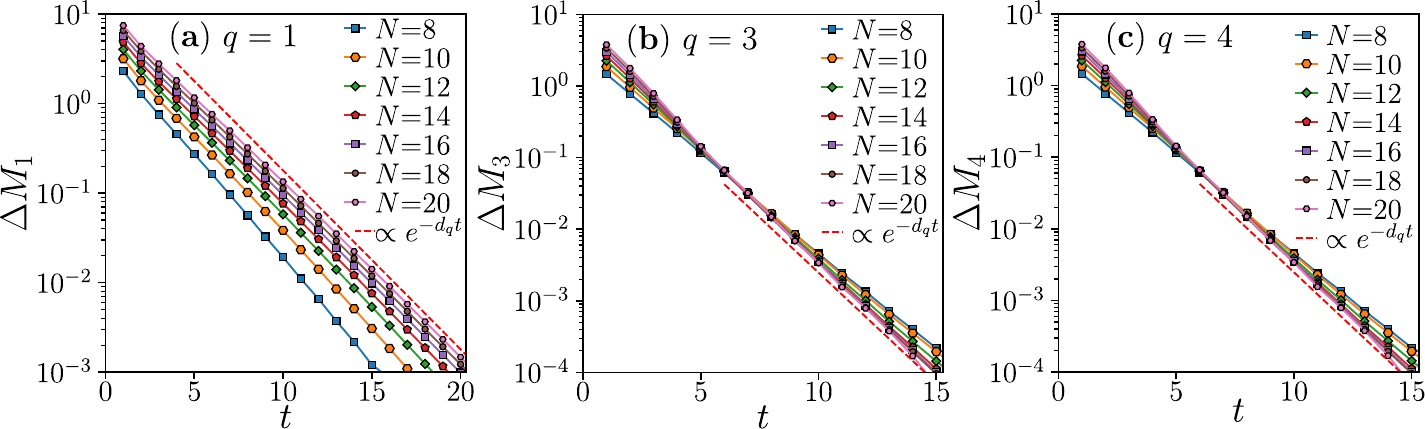}
    \caption{Dynamics under Haar-random brick-wall circuit of SRE with R\'{e}nyi indices $q=1,3,4$ for systems of $N$ qubits is shown in (a), (b), (c), respectively. The phenomenology of GSE from the Main Text is observed for $q=1$. For $q=3,4$, the SRE saturates, up to a fixed accuraccy $\epsilon$ to its long-time value, at times which are independent of $N$.
    }
    \label{fig:Q}
\end{figure}

The stabilizer R\'{e}nyi  entropy
\begin{equation}
    M_{q}(|\Psi\rangle) = \frac{1}{1-q}\log \left[\sum_{P\in\mathcal{P}_N} \frac{\langle \Psi | P |\Psi\rangle^{2q}}{d^N} \right], \label{eq:sreDEF}
\end{equation}
investigated in the Main Text for $q=2$ can be studied for arbitrary R\'{e}nyi index $q>0$, in particular, the limiting cases $q=1$ and $q=\infty$ are determined by the limits $q\to 1 $ and $q\to \infty$ of \eqref{eq:sreDEF}.
Computing SRE numerically, we observe that for $0<q\leq 2$, the phenomenology of the SRE growth remains quantitatively the same as for $q=2$ reported in the Main Text. A particular example is shown for $q=1$ in Fig.~\ref{fig:Q}(a) in which we observe the characteristic exponential decay of $\Delta M_1(t) = M^{\mathrm{Haar}}_1 - M_1(t)$ with a prefactor scaling linearly with the system size, and with saturation towards the Haar value occurring at times scaling logarithmically with the system size $N$. In contrast, for $q>2$, see Fig.~\ref{fig:Q}(b),(c) for examples of $q=3,4$, we observe that the decay of $\Delta M_q$ to a small value $\epsilon \ll 1 $ occurs at times that are independent of the system size $N$. This is consistent with observation of~\cite{turkeshi2023paulispectrummagictypical} that SRE of increasing index $q>2$ progressively fail to distinguish highly magical Haar-random states from product states with limited nonstabilizerness. Nevertheless, the saturation of $M_3$ and $M_4$ is consistent with the results reported in the Main Text in the sense that $M_2$ (and $M_{0<q<2}$), $M_Y$, and $\mathcal{M}$ saturate at the longest possible time scale $t^{\mathrm{mag}}_{\mathrm{sat}} \propto \ln(N)$ at which the higher order SRE, $M_{q>2}$, are already saturated and remain agnostic to the changes of nonstabilizerness of the state.

\end{document}